\documentclass[atmosphere,article,accept,pdftex,moreauthors]{Definitions/mdpi} 

\usepackage{amsmath}
\usepackage{graphicx}

\newcommand{\bnabla}{\boldsymbol{\nabla}}
\newcommand\bcdot{\boldsymbol{\cdot}}
\newcommand\p{\ensuremath{\partial}}
\renewcommand\Omega{\varOmega}
\renewcommand\Gamma{\varGamma}
\renewcommand\Lambda{\varLambda}
\newcommand\upi{\pi}

\def\TL#1{{\textcolor{red}{ #1}}}    % Tianyi

\def\TL#1{#1}
\renewcommand{\hl}[1]{#1}

\firstpage{1} 
\pubvolume{1}
\issuenum{1}
\articlenumber{0}
\pubyear{2023}
\copyrightyear{2023}
\externaleditor{Academic Editor: Doris Folini}
\datereceived{ 11 September 2023} 
\daterevised{12 October 2023} % Comment out if no revised date
\dateaccepted{ 20 October 2023} 
\datepublished{ } 
\hreflink{https://doi.org/} % If needed use \linebreak
\Title{Evolution of a Stratified Turbulent Cloud under Rotation}

\TitleCitation{Evolution of a Stratified Turbulent Cloud under Rotation}

\Author{\hl{Tianyi Li} %MDPI: 1. Please carefully check the accuracy of names and affiliations.2. 
$^{1,4}$, Minping Wan $^{1,2,}$* and Shiyi Chen $^{3,1,2,}$*}
\AuthorNames{Tianyi Li, Minping Wan and Shiyi Chen}

\AuthorCitation{Li, T.; Wan, M.; Chen, S.}
\address{%
$^{1}$ \quad Guangdong Provincial Key Laboratory of Turbulence Research and Applications, Department of Mechanics and Aerospace Engineering, Southern University of Science and Technology, Shenzhen 518055, China; tianyi.li@roma2.infn.it\\
$^{2}$ \quad Guangdong-Hong Kong-Macao Joint Laboratory for Data-Driven Fluid Mechanics and Engineering Applications, Southern University of Science and Technology, Shenzhen 518055, China\\
$^{3}$ \quad \hl{Eastern Institute for Advanced Study}, Ningbo 315201, China\\
$^{4}$ \quad Department of Physics and INFN, University of Rome ``Tor Vergata'', Via della Ricerca Scientifica 1,\linebreak 00133 Rome, Italy\\
}%MDPI: Please check that the address information is complete, please provide more subordinate information
\corres{Correspondence: wanmp@sustech.edu.cn (M.W.); chensy@sustech.edu.cn (S.C.)}%; Tel.: (optional; include country code; if there are multiple corresponding authors, add author initials) +xx-xxxx-xxx-xxxx (F.L.)}

\abstract{%Localized turbulence is prevalent in many geophysical flows where rotation and stratification are common features. Here we investigate the evolution of a stratified turbulent cloud under rotation. Since the turbulent cloud consists of vortices with different shapes and sizes, a single eddy under rotation and stratification is considered first. The stratified eddy is different from an eddy under pure rotation in two respects. First, a new potential vorticity (PV) mode is found, which is manifested as a static stable vortex. Second, inertial waves found in the purely rotating case are replaced by inertial-gravity waves in the stratified case. We then investigate a turbulent cloud for which we perform numerical simulations at a fixed Rossby number and for various Froude numbers. Unlike the purely rotating case where vertical columnar structures spontaneously grow out of the turbulent cloud, in the stratified case the columnar structures emerge and tilt away from the vertical direction. As the stratification gets stronger, the tilting angle increases and the rate of cloud expansion decreases. The columnar structures are composed of inertial-gravity waves, as both the tilting angles of the flow structures and vertical growth rates of the turbulent cloud can be predicted by the linear theory. We adopt Lagrangian particle tracking to identify regions dominated by wave and turbulence. When the stratification is weak, the inertial-gravity waves carry a considerable portion of energy away from the turbulent cloud. This portion of energy reduces as the stratification becomes stronger.
Localized turbulence is common in geophysical flows, where the roles of rotation and stratification are paramount. In this study, we investigate the evolution of a stratified turbulent cloud under rotation. Recognizing that a turbulent cloud is composed of vortices of varying scales and shapes, we start our investigation with a single eddy {using analytical solutions derived from a linearized system}. Compared to an eddy under pure rotation, the stratified eddy shows the physical manifestation of a known potential vorticity mode, appearing as a static stable vortex. In addition, the expected shift from inertial waves to inertial-gravity waves is observed. In our numerical simulations of the turbulent cloud, carried out at a constant Rossby number over a range of Froude numbers, stratification causes columnar structures to deviate from vertical alignment. This deviation increases with increasing stratification, slowing the expansion rate of the cloud. The observed characteristics of these columnar structures are consistent with the predictions of linear theory, particularly in their tilt angles and vertical growth rates, suggesting a significant influence of inertial-gravity waves. Using Lagrangian particle tracking, we have identified regions where wave activity dominates over turbulence. In scenarios of milder stratification, these inertial-gravity waves are responsible for a significant energy transfer away from the turbulent cloud, a phenomenon that attenuates with increasing stratification.
}

%keyword 1; keyword 2; keyword 3 (List three to ten pertinent keywords specific to the article; yet reasonably common within the subject discipline.)} 

\begin{document}

%%%%%%%%%%%%%%%%%%%%%%%%%%%%%%%%%%%%%%%%%%
%\setcounter{section}{-1} %% Remove this when starting to work on the template.
%\section{How to Use this Template}

%The template details the sections that can be used in a manuscript. Note that the order and names of article sections may differ from the requirements of the journal (e.g., the positioning of the Materials and Methods section). Please check the instructions on the authors' page of the journal to verify the correct order and names. For any questions, please contact the editorial office of the journal or support@mdpi.com. For LaTeX-related questions please contact latex@mdpi.com.%\endnote{This is an endnote.} % To use endnotes, please un-comment \printendnotes below (before References). Only journal Laws uses \footnote.

% The order of the section titles is different for some journals. Please refer to the "Instructions for Authors” on the journal homepage.

\section{Introduction}

Localized turbulence is common in both atmospheric and oceanic flows, which are strongly influenced by the Earth's rotation and stable density stratification \citep{grant1968some, nasmyth1970oceanic, wijesekera1991internal, nash2007hotspots, davidson2015turbulence, yang2021deep}. Understanding the evolution of these turbulent patches under the combined effects of rotation and stratification is crucial for gaining deeper insights into various geophysical phenomena. A key mechanism behind the generation of such turbulence is the local breakdown of internal waves \citep{riley2000fluid, sutherland2010internal}. In particular, the resulting turbulence often exhibits horizontal scales that are significantly larger than their vertical counterparts. Empirical observations from deep waters and the equatorial pycnocline also reveal pronounced differences in horizontal and vertical scales of turbulent patches \citep{nasmyth1970oceanic, wijesekera1991internal}. Furthermore, mixing induced by internal wave breaking tends to weaken local stratification. Such reduced stratification is also observed in turbulent regions within the ocean mixed layer \citep{thorpe2005turbulent}. With these characteristics in mind, our present study is concerned with a horizontally extended turbulent cloud within a vertically stably stratified, rotating fluid, where the effects of rotation {are stronger than or comparable to the} stratification. Our focus is on understanding the formation of flow structures and analyzing the associated energy dispersion patterns.

While numerous studies have highlighted the dominance of waves in spatially uniform and stationary scenarios of rotating/stratified turbulence \citep{yarom2014experimental, buzzicotti2018energy, alexakis2018cascades,di2020phase, savaro2020generation, monsalve2020quantitative}, the dynamics of localized turbulent patches has also attracted considerable attention. In particular, under rotation, these turbulent patches often give rise to columnar vortices parallel to the axis of rotation, associated with inertial waves \citep{hopfinger1982turbulence, dickinson1983oscillating, davidson2006evolution, ranjan2014evolution}. For example, in their oscillating-grid turbulence experiments, \citet{dickinson1983oscillating} found that, for a Rossby number close to unity, the boundary between perturbed and quiescent regions moves at speeds related to certain wave velocities. Similarly, \citet{davidson2006evolution} showed experimentally that the leading edges of columnar vortices under rotation align with inertial wave group velocities. Using direct numerical simulations (DNS), \citet{ranjan2014evolution} observed these columnar structures, attributed their formation to inertial waves, and emphasized their role in significant energy dissipation from localized turbulence.

Analogous to turbulent patches under rotation, the presence of internal wave radiation is clearly evident in localized turbulence within stratified fluids \citep{thorpe1982layers, browand1987behavior, de1992some, de1998experiments}, such as wakes caused by object motion \citep{gilreath1985experiments,rowe2020internal}. The flow structures in these scenarios are predominantly wave-dominated. In addition, these waves contribute significantly to energy reorganization, highlighting their fundamental role in the atmosphere and oceans \citep{maffioli2014evolution}. Anisotropic structures and wave propagation are observed in both rotating and stratified turbulent flows. These phenomena highlight the complex interplay between rotation and stratification in geophysical flows \citep{veronis1970analogy,pedlosky2013geophysical, vallis2017atmospheric}. While many studies have investigated the turbulent patch behavior in rotating and stratified fluids, the emphasis has often been on aspects such as lateral intrusions, Thorpe scales, and the vertical expansion of the patch \citep{manins1976intrusion,davies1991generation,folkard1997measurements,wells2004laboratory}. However, the waves emitted by these disturbances often receive less attention. One notable exception \citep{ranjan2016dns} used DNS to study a buoyant cloud subjected to vertical rotation and horizontal gravity, reminiscent of equatorial conditions. It was confirmed that the formation of columnar structures was due to inertial waves arising from the buoyant cloud.

Building on previous work, we explore an under-explored area: the evolution of a horizontal turbulent cloud under vertical rotation and stratification. This has significant relevance in geophysical contexts such as the high latitudes of the atmosphere and \mbox{oceans \citep{emery1984geographic}} and the deep Arctic waters \citep{jones1995deep, woodgate2001arctic}. In particular, our study examines scenarios where rotation effects outweigh stratification, in contrast to many typical geophysical flows where stratification predominates. In this context, we address three key questions: First, how does stratification affect the resulting flow structures, as opposed to purely rotating scenarios? Second, are these structures influenced by inertial-gravity waves, and how does different stratification affect their formation? Third, what fraction of the cloud's energy is transferred to waves at different stratification levels?

\section{The Evolution of a Single Stratified Eddy under Rotation}\label{sec:theoretical_analysis}

In this section, we study the evolution of a single stratified eddy under
rotation. This is motivated by the fact that a turbulent cloud can be considered as a sea of randomly oriented vortex blobs with different scales. First, we introduce {inertial-gravity} waves under the Boussinesq approximation. Second, we solve analytically the initial value problem of a compact vortex blob in a rapidly rotating stratified environment. Finally, a numerical simulation is performed to validate the analytical solution. It is shown that as the eddy evolves, part of the energy is retained in the eddy by the potential vorticity (PV) mode, while the other part is carried away from the eddy by inertial-gravity waves.

\subsection{{Inertial-Gravity} Waves}\label{subsec:igw}

The Boussinesq set of equations for a linearly stratified fluid under system rotation can be written \hl{as}:%MDPI: Please confirm all bold letter should be kept % Reply: 3. Yes. Thanks.
\begin{linenomath}
    \begin{equation}\label{equ:inc}
        \bnabla\bcdot\boldsymbol{u}=0,
    \end{equation}
    \begin{equation}
        \frac{\p\boldsymbol{u}}{\p t}+\boldsymbol{u}\bcdot\bnabla\boldsymbol{u}=-\frac{1}{\rho_0}\bnabla p+2\boldsymbol{u}\times\boldsymbol{\Omega}-N\phi\boldsymbol{e}_z+\nu\nabla^2\boldsymbol{u},
    \end{equation}
    \begin{equation}\label{equ:den}
        \frac{\p\phi}{\p t}+\boldsymbol{u}\bcdot\bnabla\phi=Nu_z+\kappa\nabla^2\phi,
    \end{equation}
\end{linenomath}
where $\boldsymbol{u}$ is the velocity vector, $u_z$ is the $z$-component of the velocity, $p$ is the modified pressure incorporating a centrifugal term, $\boldsymbol{\Omega}=\Omega\boldsymbol{e}_z$ represents the rotation vector, $N$ is the Brunt--V\"ais\"al\"a frequency, and $\nu$ and $\kappa$ are the kinematic viscosity and the diffusion coefficient, respectively. We define $\phi=\left(g/\rho_0N\right)\rho'$, which has the dimension of the velocity, and $\rho'$ is the density perturbation from the ambient density.

Inertial-gravity waves can be obtained after a linearization of the Boussinesq equations. {Neglecting the molecular diffusion and the second-order terms of $\boldsymbol{u}$, $\boldsymbol{\omega}=\bnabla\times\boldsymbol{u}$ and $\phi$ one obtains:}
\begin{linenomath}
    \begin{equation}\label{equ:lvor}
        \frac{\p\boldsymbol{\omega}}{\p t}+\bnabla\times(2\boldsymbol{\Omega}\times\boldsymbol{u})=-N\bnabla\phi\times\boldsymbol{e}_z,
    \end{equation}
    \begin{equation}\label{equ:lphi}
        \frac{\p\phi}{\p t}=Nu_z,
    \end{equation}
\end{linenomath}
from which one finds (see Ref. \citep{lesieur1987turbulence}, p. 56)
\begin{linenomath}
    \begin{equation}\label{equ:luz}
        \frac{\p^2}{\p t^2}\nabla^2u_z+4\Omega^2\frac{\p^2u_z}{\p z^2}+N^2\left(\frac{\p^2u_z}{\p x^2}+\frac{\p^2u_z}{\p y^2}\right)=0.
    \end{equation}
\end{linenomath}

This equation admits inertial-gravity waves with the dispersion relation given as
\begin{linenomath}
    \begin{equation}\label{equ:fig}
        \varpi=\pm\sqrt{N^2k_h^2+4\Omega^2k_z^2}/k,
    \end{equation}
\end{linenomath}
where $\varpi$ is the frequency, while $k_h=\sqrt{k_x^2+k_y^2}$ and $k=\sqrt{k_x^2+k_y^2+k_z^2}$ are the horizontal wavenumber and the total wavenumber, respectively. The group velocity is $\boldsymbol{c}_g=\bnabla\varpi$, whose $z$-component is
\begin{linenomath}
    \begin{equation}
        c_{g,z}=\frac{\p\varpi}{\p k_z}=\pm\frac{\left(4\Omega^2-N^2\right)k_h^2k_z}{\sqrt{N^2k_h^2+4\Omega^2k_z^2}k^3}.
    \end{equation}
\end{linenomath}

\subsection{Analytical Study of a Single Eddy}\label{subsec:aac}

Following \citet{davidson2006evolution}, we consider the evolution of an eddy in a stratified fluid under rotation, assuming that the process is axisymmetric with respect to the eddy axis. In this context, in a cylindrical coordinate system $(r,\theta,z)$, an axisymmetric velocity field $\boldsymbol{u}$ can be decomposed into azimuthal and poloidal components
\begin{linenomath}
\begin{equation}\label{equ:avf}
\boldsymbol{u}=\left(\Gamma/r\right)\boldsymbol{e}_\theta+\bnabla\times\left[\left(\psi/r\right)\boldsymbol{e}_\theta\right],
\end{equation}
\end{linenomath}
where $\Gamma=u_\theta r$ is the angular momentum and $\psi$ is the Stokes streamfunction. We substitute Equation (\ref{equ:avf}) into the linearized inviscid vorticity Equation (\ref{equ:lvor}) and the linearized non-diffusive Equation (\ref{equ:lphi}) to obtain
\begin{linenomath}
    \begin{equation}\label{equ:lin}
        \frac{\p \Gamma}{\p t}=2\Omega\frac{\p\psi}{\p z},\quad\frac{\p}{\p t}\left(r\omega_\theta\right)=2\Omega\frac{\p\Gamma}{\p z}+Nr\frac{\p\phi}{\p r},\quad\frac{\p\phi}{\p t}=\frac{N}{r}\frac{\p\psi}{\p r},
    \end{equation}
\end{linenomath}
where $\nabla_*^2\psi=\left(r\p/\p r\right)\left(r^{-1}\p\psi/\p r\right)+\p^2\psi/\p z^2=-r\omega_\theta$, $\nabla_*^2$ being the Stokes operator. Combining the equations in (\ref{equ:lin}) yields
\begin{linenomath}
    \begin{equation}\label{equ:cmb}
        \frac{\p}{\p t}\left(\frac{\p^2}{\p t^2}\nabla_*^2\Gamma+\left[\left(2\Omega\right)^2\frac{\p^2}{\p z^2}+N^2\left(r\frac{\p}{\p r}\right)\left(\frac{\p}{r\p r}\right)\right]\Gamma\right)=0.
    \end{equation}
\end{linenomath}
Equation (\ref{equ:cmb}) can be readily solved by using the Hankel-cosine transform
\begin{linenomath}
    \begin{equation}
        \hat{u}_\theta=\frac{1}{2\upi^2}\int_0^\infty\int_0^{\infty}ru_\theta J_1\left(k_rr\right)\cos\left(k_zz\right)\mathrm{d}r\mathrm{d}z,
    \end{equation}
\end{linenomath}
where $J_1$ is the Bessel function of the first kind of order $1$. Here, $k_r$ and $k_z$ represent the wavenumbers in the $r$-direction and the $z$-direction, respectively, with $0\le k_r,k_z<\infty$. Given the initial conditions $\hat{u}_\theta=\hat{u}_\theta^{\left(0\right)}$, $\psi=0$ and $\phi=0$, we can derive that
\begin{linenomath}
\begin{equation}\label{equ:lss}
\hat{u}_\theta=\left[\hat{u}_\theta^{\left(0\right)}\!/\!\left(N^2k_r^2\!+\!\left(2\Omega\right)^2k_z^2\right)\right]\left[N^2k_r^2\!+\!\left(2\Omega\right)^2k_z^2\cos\left(\sqrt{N^2k_r^2\!+\!\left(2\Omega\right)^2k_z^2}t/k\right)\right],
\end{equation}
\end{linenomath}
from which we obtain
\begin{linenomath}
% \begin{align}\label{equ:lsp}
% u_\theta=\,&4\upi\int_0^\infty\int_0^\infty k_r\left[N^2k_r^2\hat{u}_\theta^{\left(0\right)}\!/\!\left(N^2k_r^2\!+\!\left(2\Omega\right)^2k_z^2\right)\right]J_1\left(k_rr\right)\cos\left(k_zz\right)\mathrm{d}k_r\mathrm{d}k_z\nonumber\\
% &+2\upi\int_0^\infty\int_0^\infty k_r\left[\left(2\Omega\right)^2k_z^2\hat{u}_\theta^{\left(0\right)}\!/\!\left(N^2k_r^2\!+\!\left(2\Omega\right)^2k_z^2\right)\right]J_1\left(k_rr\right)\bigg[\cos\bigg(k_zz\nonumber\\
% &\left.\left.-\sqrt{N^2k_r^2\!+\!\left(2\Omega\right)^2k_z^2}t/k\right)\!+\!\cos\left(k_zz\!+\!\sqrt{N^2k_r^2\!+\!\left(2\Omega\right)^2k_z^2}t/k\right)\right]\mathrm{d}k_r\mathrm{d}k_z.
% \end{align}
\vspace{-6pt}
\begin{adjustwidth}{-\extralength}{0cm}
\begin{align}\label{equ:lsp}
u_\theta=\,&\underbrace{4\upi\int_0^\infty\int_0^\infty \left(N^2k_r^3\hat{u}_\theta^{\left(0\right)}\!/\!\alpha^2\right)J_1\left(k_rr\right)\cos\left(k_zz\right)\mathrm{d}k_r\mathrm{d}k_z}_{\text{PV mode}}\nonumber\\
&+\underbrace{2\upi\int_0^\infty\int_0^\infty \left[\left(2\Omega\right)^2k_rk_z^2\hat{u}_\theta^{\left(0\right)}\!/\!\alpha^2\right]J_1\left(k_rr\right)\left[\cos\left(k_zz\!-\!\alpha t/k\right)\!+\!\cos\left(k_zz\!+\!\alpha t/k\right)\right]\mathrm{d}k_r\mathrm{d}k_z}_{\text{inertial-gravity waves}},
\end{align}
\end{adjustwidth}
\end{linenomath}
where we introduce $\alpha=\sqrt{N^2k_r^2\!+\!\left(2\Omega\right)^2k_z^2}$ for conciseness. In the absence of stratification (i.e., when $N=0$), Equations (\ref{equ:lss}) and (\ref{equ:lsp}) are consistent with the results presented in \citet{davidson2006evolution}, which consider only the rotation of the system. However, the inclusion of stratification introduces a new time-independent term and changes the frequencies of the previously identified wave-like components. Note that the phase velocities of the wave-like components, given by $\pm\alpha/k$, closely match those of the inertial-gravity waves according to Equation (\ref{equ:fig}), provided we replace $k_r$ with $k_h$. This alignment strongly suggests that these wave-like components are indeed manifestations of the inertial-gravity waves. {In a spatially uniform rotating stratified flow with periodic boundary condition}, \citet{smith2002generation} identified the linear eigenmodes as two inertial-gravity waves and a PV mode, suggesting that the time-independent term in (\ref{equ:lsp}) is consistent with the zero-frequency PV mode. However, our solution shows how these PV mode and inertial-gravity waves manifest in {the flow evolution from a spatially localized initial condition}---an aspect not addressed in previous studies. Later in Section \ref{subsec:nvaa}, we will explore the associated flow structures and elucidate their role in energy dispersion.

\subsection{Numerical Validation of the Analytical Results}\label{subsec:nvaa}

To assess the applicability of the analytical results in Section \ref{subsec:aac} and to develop a better understanding of the evolution of a single eddy in a rotating stratified fluid, a simple initial velocity field of the Gaussian-eddy form is chosen:
\begin{linenomath}
    \begin{equation}\label{equ:Gev}
        \boldsymbol{u}=\Lambda r\exp\left(-\frac{r^2+z^2}{\delta^2}\right)\boldsymbol{e}_\theta,
    \end{equation}
\end{linenomath}
for which \citet{davidson2006evolution} have given an analytical solution when only the rotation is considered. In this equation, $\Lambda$ is the characteristic angular rotation rate and $\delta$ is the characteristic size. The corresponding Rossby and Froude numbers are defined by \mbox{$Ro=\Lambda/(2\Omega)$} and $Fr=\Lambda/N$, respectively. To validate our analytical solutions, we chose parameters $Ro=0.02$ and $Fr=0.16$, corresponding to a vortex characterized by $\Lambda=0.422$ and $\delta=0.125$. {This choice corresponds to a parameter set from our turbulent cloud studies in Section \ref{sec:rot_str_turb_cloud}. We note that the validation of our analytical solutions remains robust across different parameter sets in these studies.} The analytical solution is obtained by substituting (\ref{equ:Gev}) into (\ref{equ:lsp}). We then performed a DNS to study the evolution of this Gaussian eddy. Equations (\ref{equ:inc})--(\ref{equ:den}) are solved using a parallelized pseudo-spectral code in a $512^3$ periodic box of size $l_{box}=2\upi$, where the fourth-order Runge--Kutta time-stepping scheme is \mbox{employed {\citep{li2020flow}}}. The linear terms caused by rotation and stratification, together with the viscous and diffusive terms, are integrated exactly by using an integrating factor technique. A combination of phase-shifting and truncation is used to de-alias the nonlinear terms (see e.g., Ref. \citep{canuto2012spectral}).

\begin{figure}[H]
	\begin{minipage}[b]{.5\textwidth}
        \includegraphics[trim=2pt 2pt 2pt 2pt, clip, width=0.7\textwidth]{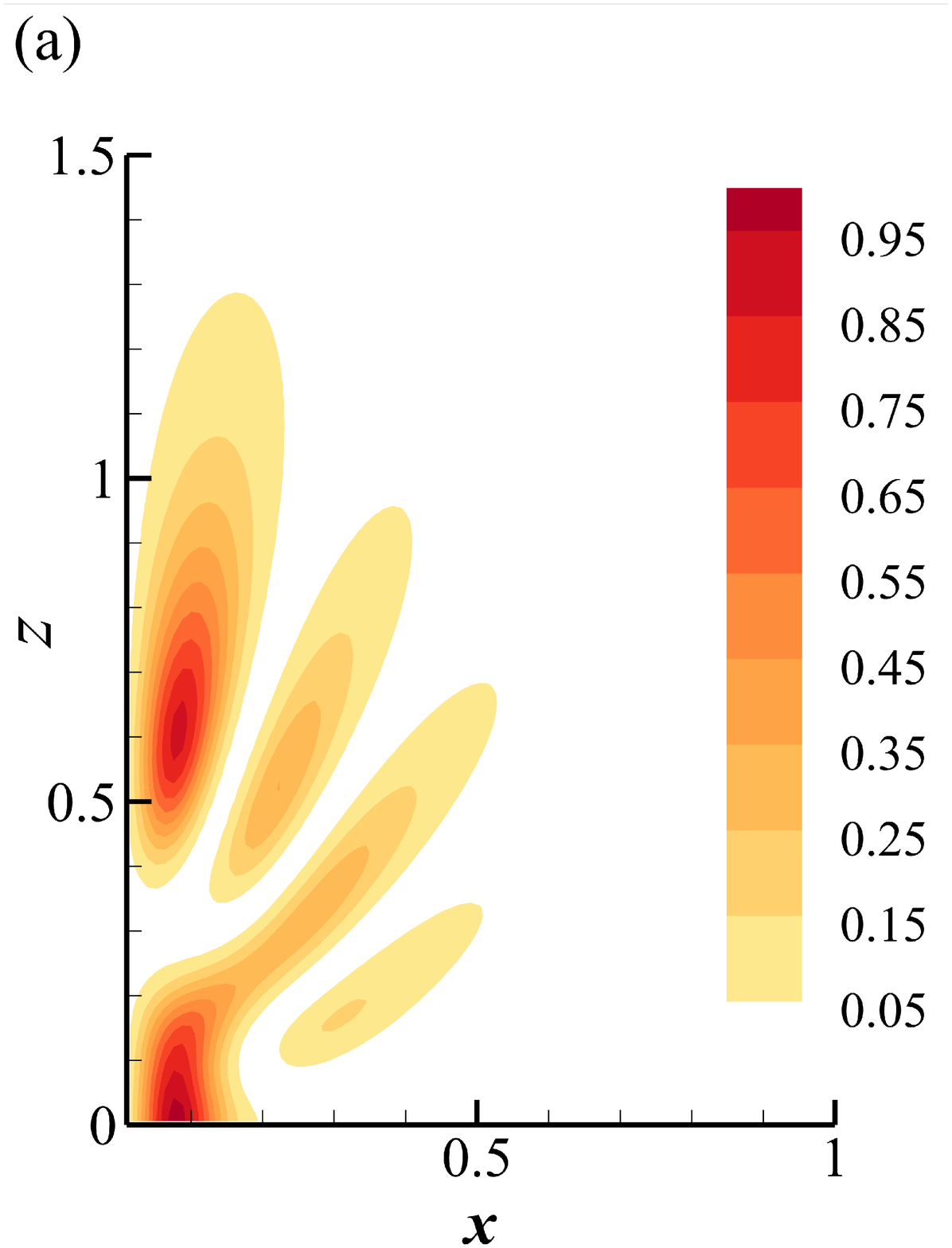}
	\end{minipage}
	\hfill
	\begin{minipage}[b]{.5\textwidth}
		\includegraphics[trim=2pt 2pt 2pt 2pt, clip, width=0.7\textwidth]{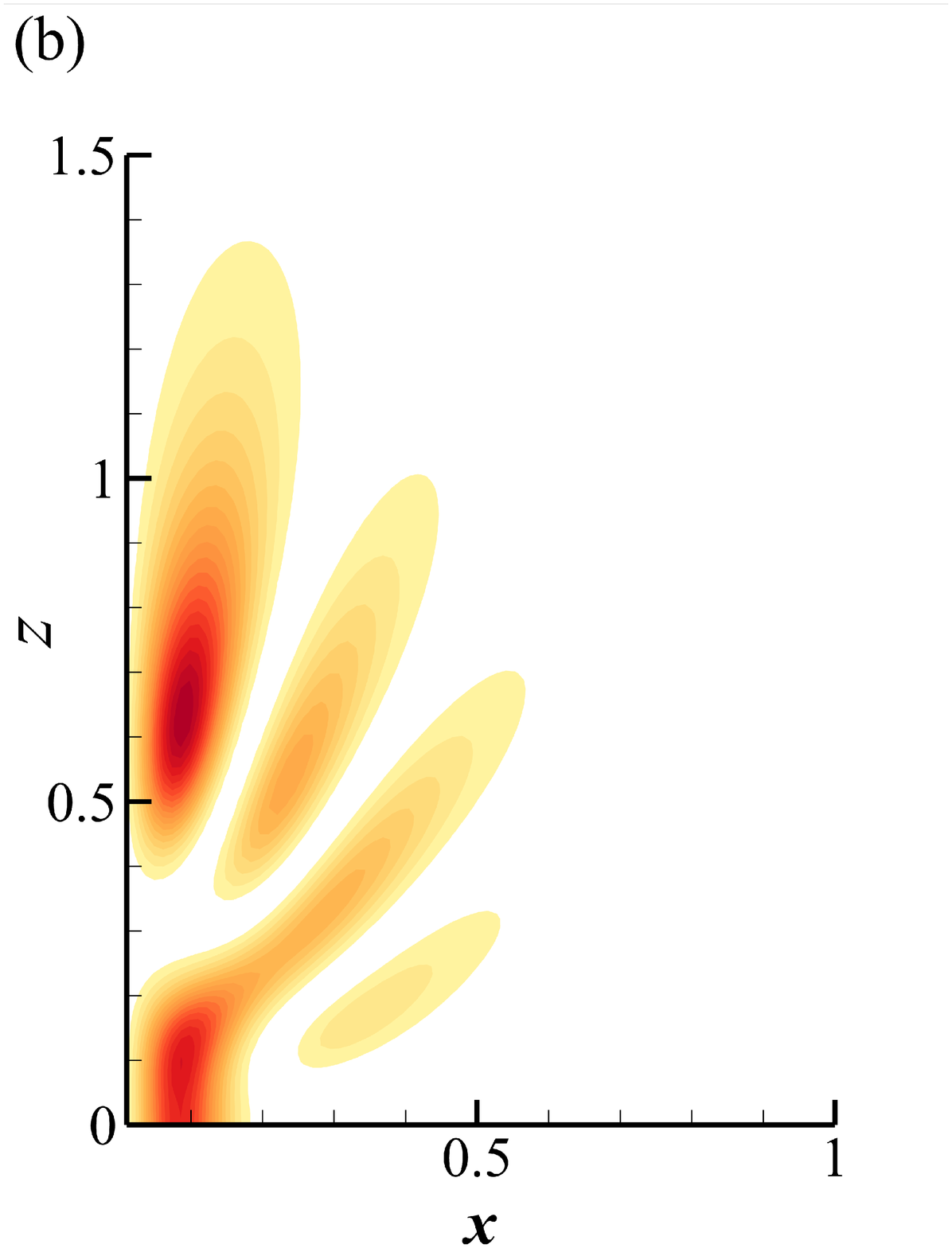}
	\end{minipage}
	\caption{\hl{Contours} of $u_\theta^2/(u_\theta^2)_{max}$ in the \emph{x}-\emph{z} plane through the axis of the Gaussian eddy at $t/t_f=13.5$. The results are from (\textbf{a}) the analytical solution (\ref{equ:lsp}) and (\textbf{b}) the DNS.}\label{fig:validation}%MDPI: Please bold all subfigure number in all main text % Reply: 5. Fixed. Thanks.
\end{figure}

Figures \ref{fig:validation}\TL{\textbf{a},\textbf{b}} show the contours of $u_\theta^2$ (normalized by its maximum value) in the \emph{x}-\emph{z} plane passing through the axis of the Gaussian eddy at $t/t_f=13.5$. These are results from the analytical solution and the DNS, respectively. Here $t_f=1/(2\Omega)$ denotes the rotation time scale. {The DNS result, which considers the full Navier--Stokes equations including viscous and nonlinear effects, is still in close agreement with the analytical counterpart. Although there are discrepancies due to these effects, the overall trend and structure are consistent between the two.} A prominent region of $u_\theta^2$ near the origin stands out, which is absent in the unstratified scenario (Ref. \TL{\citep{ranjan2014evolution},}  \hl{Figure 3}). %MDPI: Please check if this Figure 3 belongs to ref, not this manuscript. Please make sure the citations appear in sequential order. % Reply: 4. fixed. Thanks
This region is indicative of the PV mode, while the other intense regions represent inertial-gravity waves.

\begin{figure}[H]
	\begin{minipage}[b]{.5\textwidth}
		\includegraphics[trim=2pt 2pt 2pt 2pt, clip, width=0.55\textwidth]{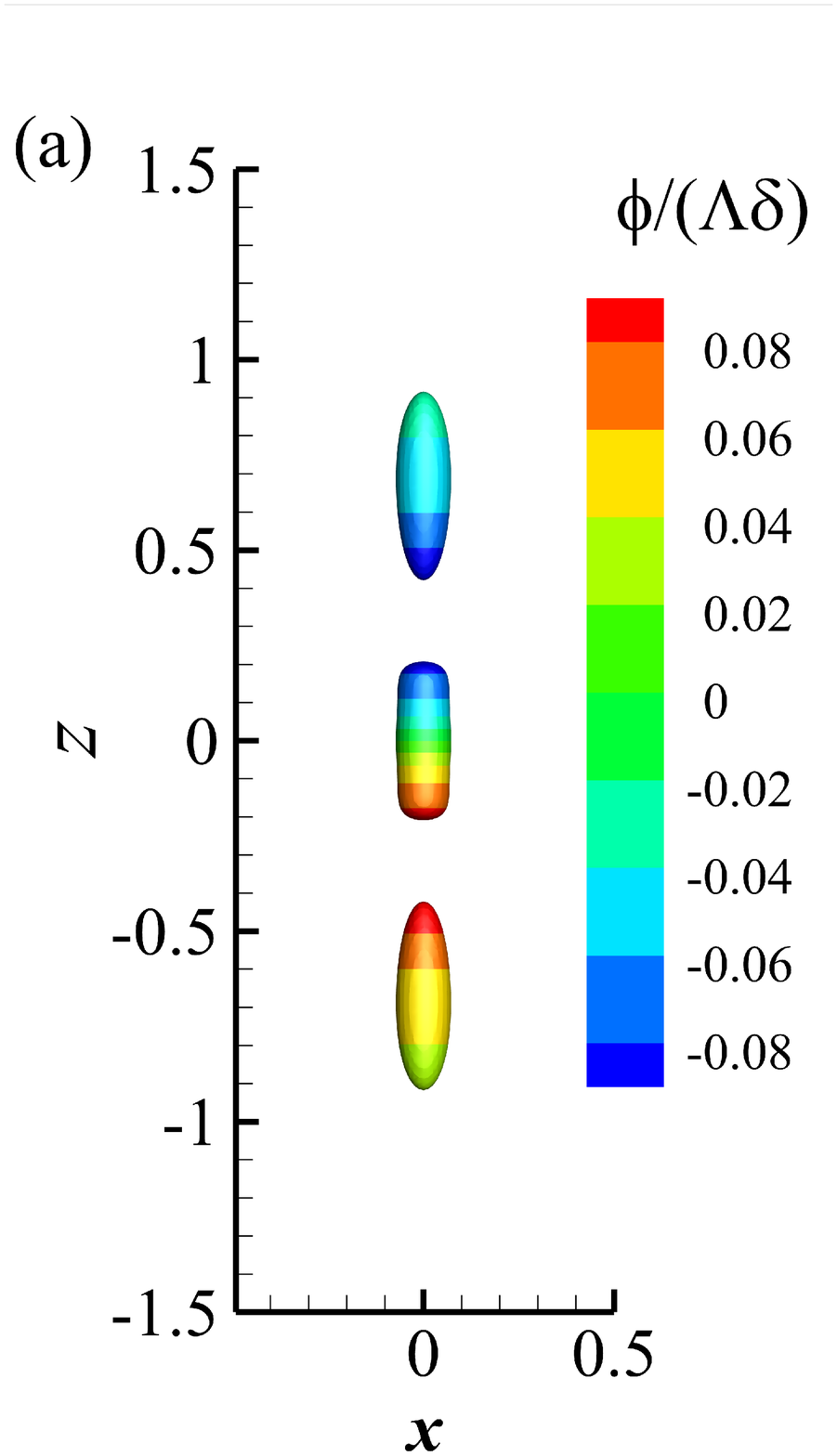}
	\end{minipage}
	\hfill
	\begin{minipage}[b]{.5\textwidth}
		\includegraphics[trim=2pt 2pt 2pt 2pt, clip, width=0.55\textwidth]{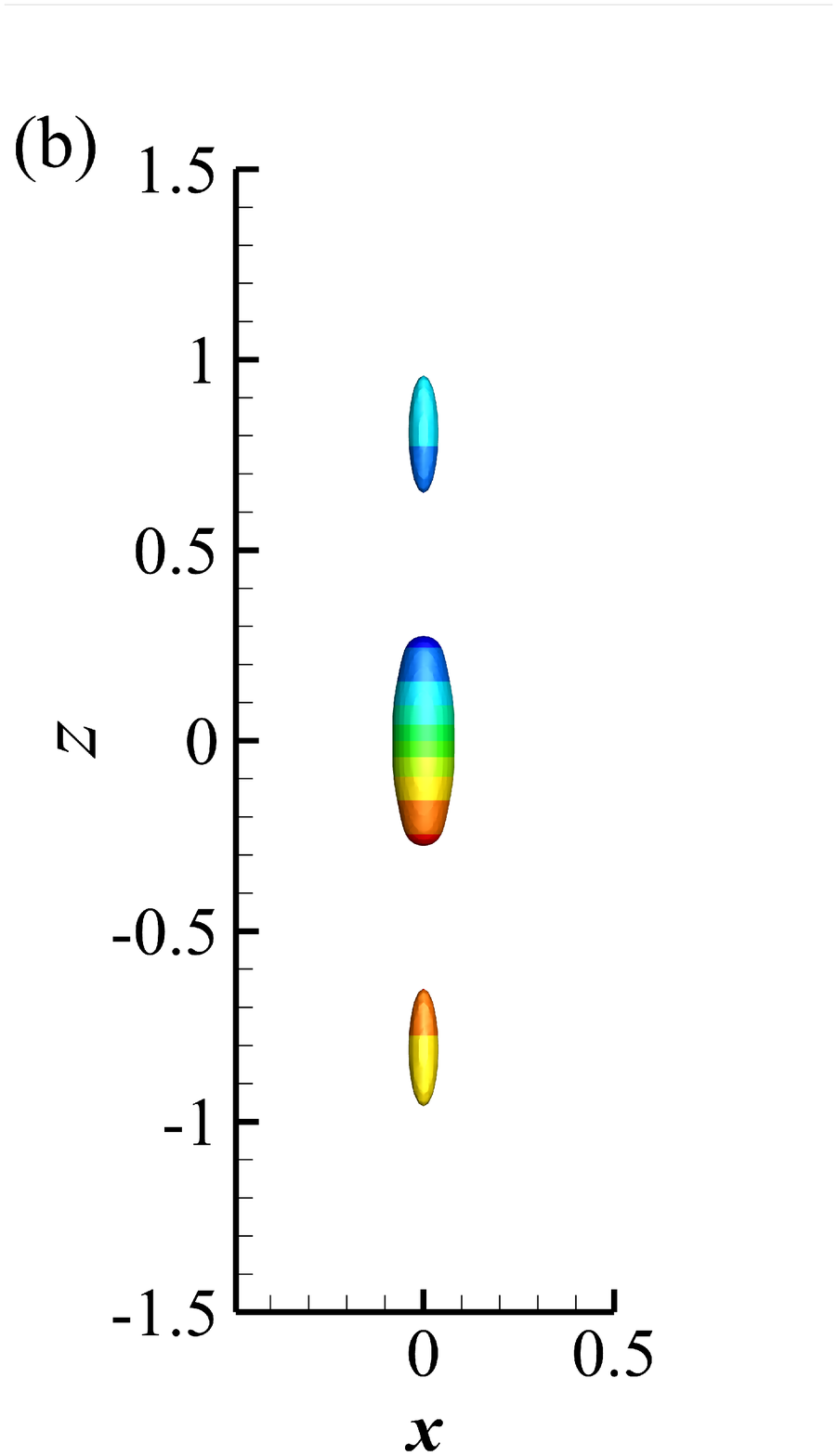}
	\end{minipage}
	\caption{\hl{The iso-surfaces} %MDPI: Please change the hyphen (-) into a minus sign ($-$, "U+2212"). e.g., "-1" should be "$-$1". % Reply: 6. Sorry, I do not understand which "hypen"s. Do you mean those in "iso-surfaces", "inertial-gravity", "under-explored", ...? It would look strange if I do that, Please be more specific about which words. Thanks
 of $\omega_z=0.1\Lambda$ (front view) color-coded with $\phi$ obtained from the DNS at (\textbf{a}) $t/t_f=13.5$ and (\textbf{b}) $t/t_f=16.9$.}\label{fig:wzcc}
\end{figure}

To illustrate the evolution of the flow structures and their corresponding energy dispersion, Figure \ref{fig:wzcc} shows iso-surfaces of the vertical vorticity, $\omega_z=0.1\Lambda$, color-coded with $\phi$, at times $t/t_f=13.5$ and $16.9$. As the system begins to evolve, a central vortex appears at the origin, flanked by two vortices, one above and one below (Figure \ref{fig:wzcc}\TL{\textbf{a}}). Over time, the central vortex undergoes minor changes, while the flanking vortices become smaller and their centers move further away from the origin (Figure \ref{fig:wzcc}\TL{\textbf{b}}). This behavior is in stark contrast to the purely rotating case \citep{ranjan2016dns}, where only two columnar structures emerge, progressively moving away from each other as they elongate. The shifts in the flow structures can be attributed to the interplay between the Coriolis and buoyancy forces. In Figure \ref{fig:wzcc}\TL{\textbf{a}}, all three cyclonic vortices ($w_z>0$) are influenced by the Coriolis force, causing them to stretch horizontally. For the central vortex, $\phi$ has negative values at the top and positive ones at the base. {This distribution results in a vertical stretch from the buoyancy force, which causes a horizontal contraction of the vertex due to the incompressibility of the fluid, counteracting the Coriolis force.} As a result, this vortex remains relatively stable, maintaining a nearly consistent energy distribution over time. For each flanking vortex, $\phi$ retains its sign, showing larger absolute values for smaller magnitudes of $|z|$. This implies that the buoyancy force is acting to move these vortices away from the origin and to compress them vertically. The interaction of the Coriolis and buoyancy forces causes these vortices to recede and decrease in size, resulting in energy dispersion and potential-kinetic energy exchange. It is worth noting that a similar phenomenon to the central vortex has been documented in {fully developed} rotating stratified turbulence, where it correlates with a pronounced exchange between kinetic and potential energy \citep{li2020flow}.

\vspace{-20pt}

\section{A Stratified Turbulent Cloud under Rotation}\label{sec:rot_str_turb_cloud}

In this section, we move from the study of a single eddy to the nonlinear evolution of a turbulent cloud. {We observe characteristics compatible with inertial-gravity waves in DNS by studying the flow structures.} Finally, we calculate the energy {exchanged} from the turbulent cloud by these waves to understand their role in energy transfer for localized turbulent patches.

\subsection{DNS of a Turbulent Cloud}

\begin{figure}[H]
	\begin{minipage}[b]{1.0\textwidth}
		\centering
		\includegraphics[trim=2pt 2pt 2pt 2pt, clip, width=0.329\textwidth]{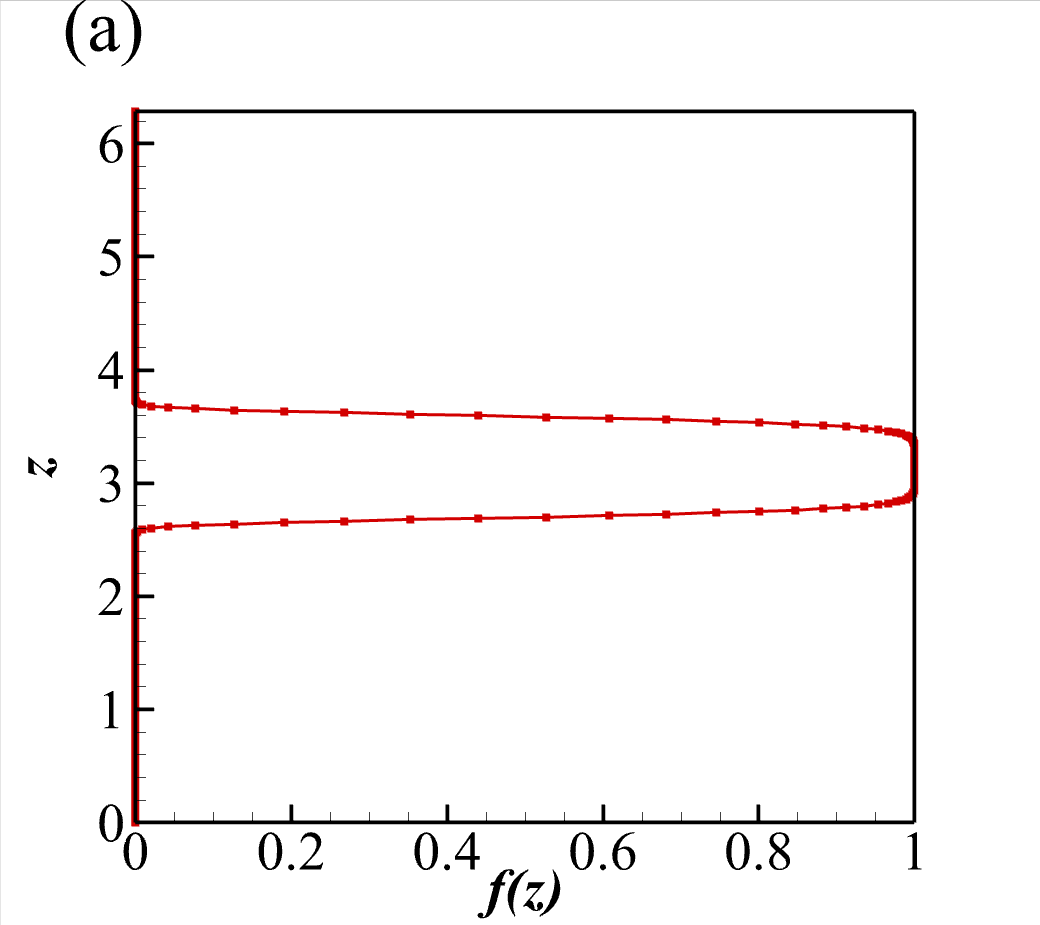}
		\includegraphics[trim=2pt 2pt 2pt 2pt, clip, width=0.329\textwidth]{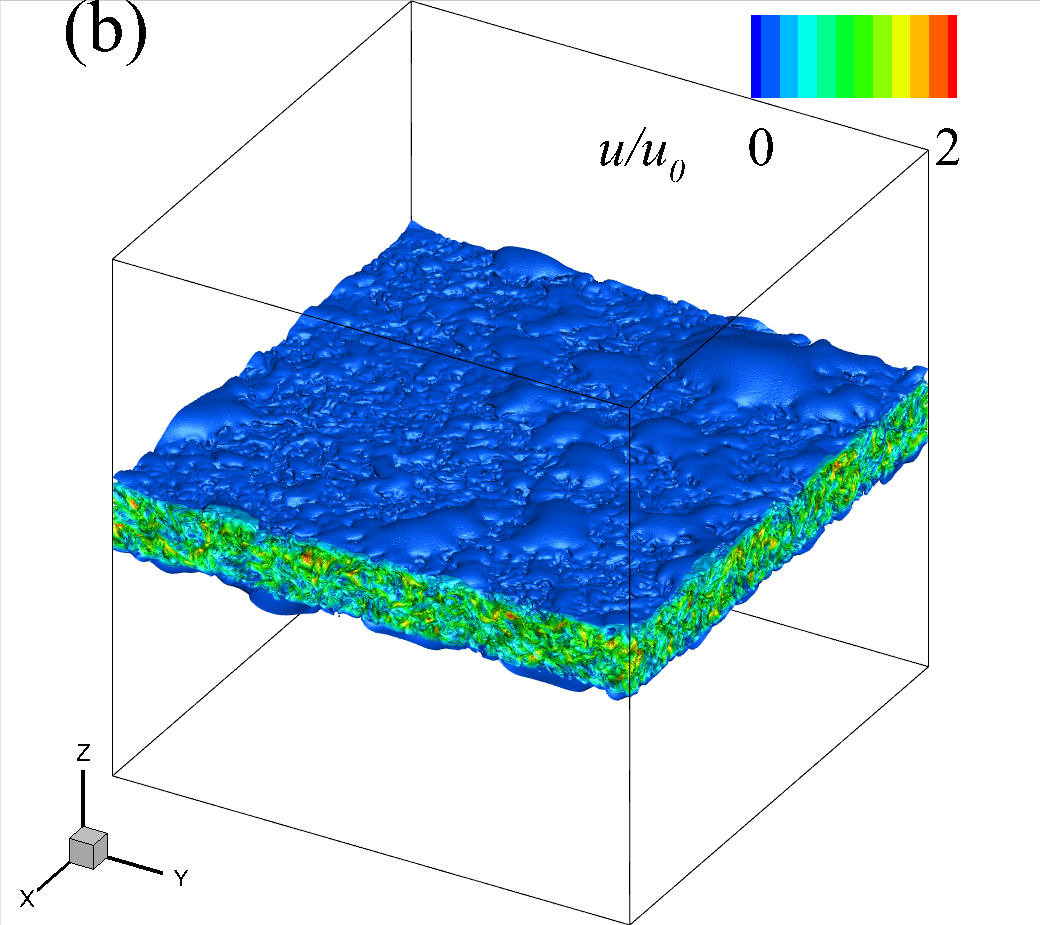}
		\includegraphics[trim=2pt 2pt 2pt 2pt, clip, width=0.329\textwidth]{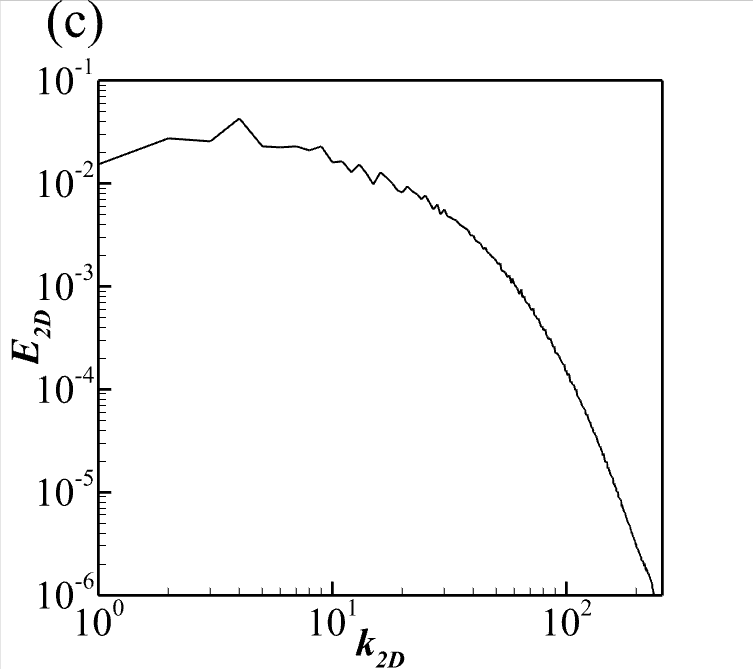}
	\end{minipage}
	\caption{(\textbf{a}) \hl{The} %MDPI: Please change the hyphen (-) into a minus sign ($-$, "U+2212"). e.g., "-1" should be "$-$1". % Reply: 7. Sorry, I do not understand which "hypen"s. Do you mean those in "iso-surfaces", "inertial-gravity", "under-explored", ...? It would look strange if I do that, Please be more specific about which words. Thanks
 spatial filter given by $f\left(z\right)=\exp\left(-10^a\left|z-\upi\right|^m\right)$ with $a=4$ and $m=9$. (\textbf{b}) Initial turbulent cloud visualized by the velocity module. (\textbf{c}) Energy spectrum in the $x$-$y$ plane, averaged over different $z$ values around the center of the cloud.}\label{fig:spat_filter}
\end{figure}

Following the procedure described by {\citet{ranjan2014evolution}}, we performed DNS on fully developed homogeneous isotropic turbulence to obtain the initial velocity field for the turbulent cloud study. We then selected a velocity field corresponding to a specific moment characterized by an integral length scale of \mbox{{$l_0\sim\int_{0}^{\infty}k^{-1}E(k)\mathrm{d}k/\int_{0}^{\infty}E(k)\mathrm{d}k\approx0.04l_{box}$}} and a Reynolds number $Re_0=u_0l_0/\nu\approx169$, where {$E(k)$ is the energy spectrum and} $u_0$ is the r.m.s. velocity. To numerically generate a horizontal turbulent cloud, the resulting velocity field was spatially filtered with
\begin{linenomath}
    \begin{equation}
        f\left(z\right)=\exp\left(-10^a\left|z-\upi\right|^m\right),
    \end{equation}
\end{linenomath}
as shown in Figure \ref{fig:spat_filter}\TL{\textbf{a}}. Here we set $a=4$ and $m=9$, resulting in a cloud thickness of $l_c=0.13l_{box}$, approximately three times the integral length scale $l_0$. {The velocity field is then projected to be divergence-free.} The buoyancy field was initialized with $\phi=0$ throughout the computational domain, which ensures a uniformly linear density profile, indicating no initial mixing within the cloud. Figure \ref{fig:spat_filter}\TL{\textbf{b}} shows the initial turbulent cloud as indicated by the velocity module, and its spectrum in the $x$-$y$ plane, averaged over its central part in the $z$-direction, is shown in Figure \ref{fig:spat_filter}\TL{\textbf{c}}.

To investigate the effect of {the} stratification on the evolution of a turbulent cloud under rotation, we conducted DNS with a constant Rossby number, $Ro=0.11$, and five different Froude numbers: $Fr=\infty$ (corresponding to $N=0$, a purely rotating scenario), $0.88$, $0.44$, $0.22$ and $0.11$. {All simulations started from the same initial condition with no external forcing (i.e., the decaying case).} The simulation parameters are given in \mbox{Table \ref{tab:spd}}, where $Ro=u_0/(2\Omega l_0)$ and $Fr=u_0/(Nl_0)$. With the exception of the purely rotating case R0.11, the simulations are denoted by their Froude numbers, e.g., the case F0.88 corresponds to $Fr=0.88$. {For practical reference, we have also listed the Burger number, \mbox{$Bu=(Ro/Fr)^2$ \citep{cushman2011introduction}}, for each case in Table \ref{tab:spd}. This number indicates the balance between stratification and rotation in the flows studied.}
\begin{table}[H]
	\begin{center}	\caption{Simulation parameters for the DNS of a turbulent cloud. {In all cases, we have set $\nu=\kappa=0.001$.}}
\setlength{\cellWidtha}{\textwidth/10-2\tabcolsep-0.15in}
\setlength{\cellWidthb}{\textwidth/10-2\tabcolsep+0.45in}
\setlength{\cellWidthc}{\textwidth/10-2\tabcolsep-0.2in}
\setlength{\cellWidthd}{\textwidth/10-2\tabcolsep-0.1in}
\setlength{\cellWidthe}{\textwidth/10-2\tabcolsep+0in}
\setlength{\cellWidthf}{\textwidth/10-2\tabcolsep-0in}
\setlength{\cellWidthg}{\textwidth/10-2\tabcolsep-0in}
\setlength{\cellWidthh}{\textwidth/10-2\tabcolsep-0in}
\setlength{\cellWidthi}{\textwidth/10-2\tabcolsep+0in}
\setlength{\cellWidthj}{\textwidth/10-2\tabcolsep+0in}
\scalebox{1}[1]{\begin{tabularx}{\textwidth}{>{\centering\arraybackslash}m{\cellWidtha}>{\centering\arraybackslash}m{\cellWidthb}>{\centering\arraybackslash}m{\cellWidthc}>{\centering\arraybackslash}m{\cellWidthd}>{\centering\arraybackslash}m{\cellWidthe}>{\centering\arraybackslash}m{\cellWidthf}>{\centering\arraybackslash}m{\cellWidthg}>{\centering\arraybackslash}m{\cellWidthh}>{\centering\arraybackslash}m{\cellWidthi}>{\centering\arraybackslash}m{\cellWidthj}}

            \toprule
			\textbf{Case}  & \textbf{Resolution} & \boldmath$Ro$ & \boldmath$2\Omega$ & \boldmath$Fr$     & \boldmath$N$  & \boldmath{$Bu$} & \boldmath{$Re_0$} & \boldmath$l_c/l_{box}$ & \boldmath$l_c/l_0$ \\
            \midrule
			R0.11 & $512^3$    & 0.11 & 21.1      & $\infty$ & ~0.0 & {0.00} & {169}  & 0.13          & 3\\
			F0.88 & $512^3$    & 0.11 & 21.1      & 0.88     & ~2.6 & {0.02} & {169}  & 0.13          & 3\\
			F0.44 & $512^3$    & 0.11 & 21.1      & 0.44     & ~5.3 & {0.06} & {169}  & 0.13          & 3\\
			F0.22 & $512^3$    & 0.11 & 21.1      & 0.22     & 10.6 & {0.25} & {169}  & 0.13          & 3\\
			F0.11 & $512^3$    & 0.11 & 21.1      & 0.11     & 21.1 & {1.00} & {169}  & 0.13          & 3\\
            \bottomrule
		\end{tabularx}}
	
		\label{tab:spd}
	\end{center}
\end{table}

\vspace{-20pt}
\subsection{Evolution of Flow Structures}\label{subsec:evo_flow_struct}

\begin{figure}[H]
	\begin{minipage}[b]{1.0\textwidth}
		\centering
		\includegraphics[trim=2pt 2pt 2pt 2pt, clip, width=.3\textwidth]{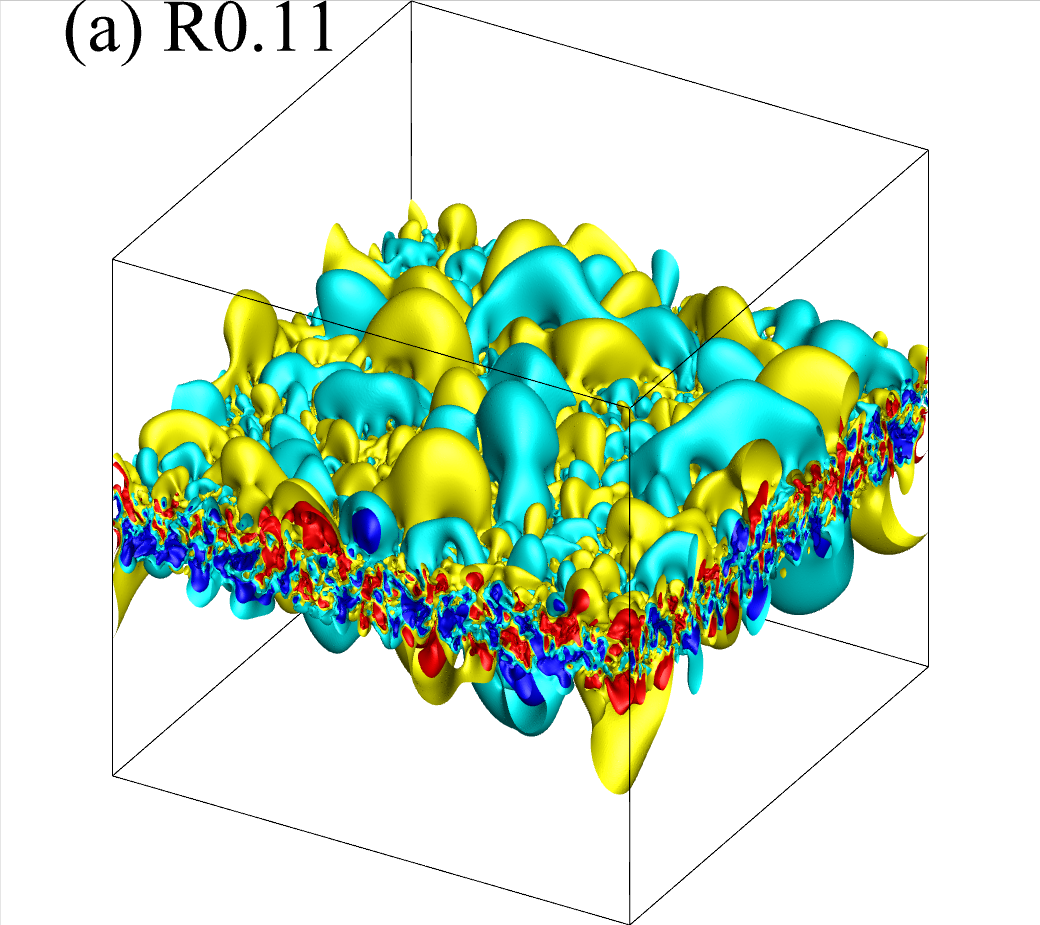}
		\includegraphics[trim=2pt 2pt 2pt 2pt, clip, width=.3\textwidth]{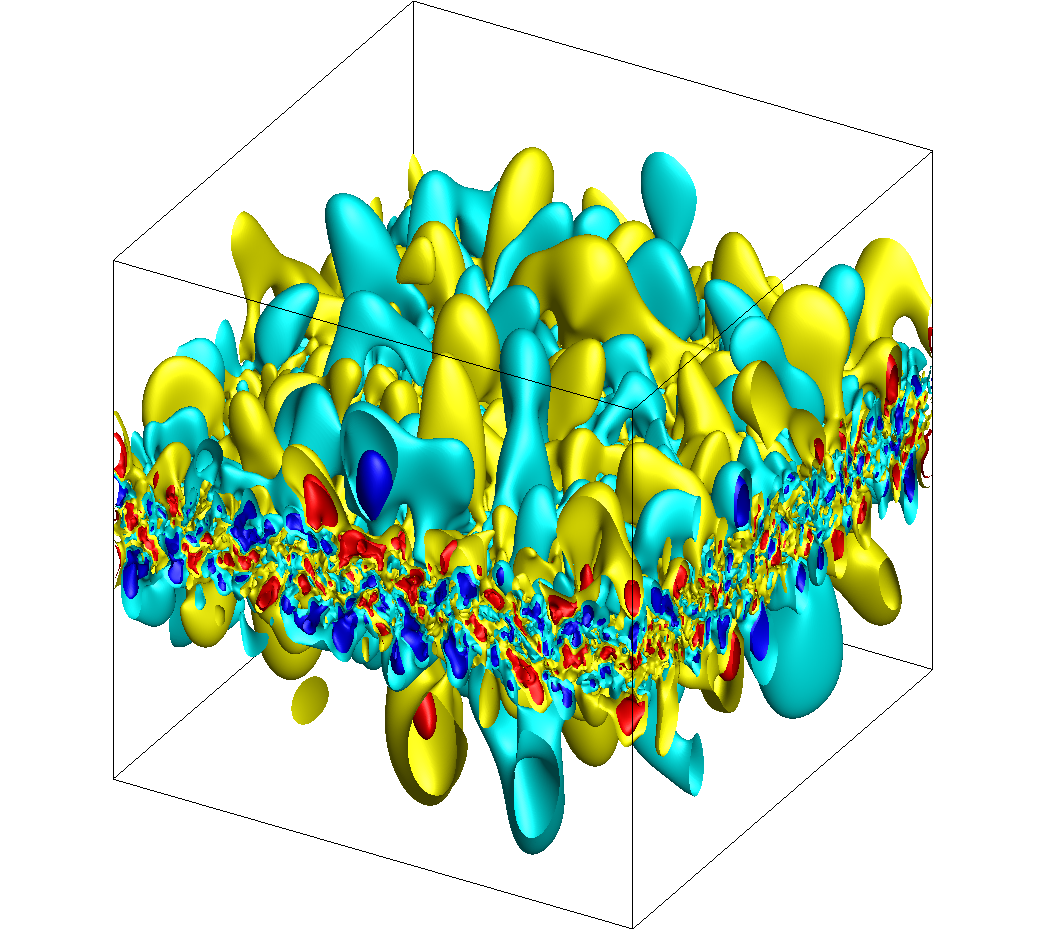}
		\includegraphics[trim=2pt 2pt 2pt 2pt, clip, width=.3\textwidth]{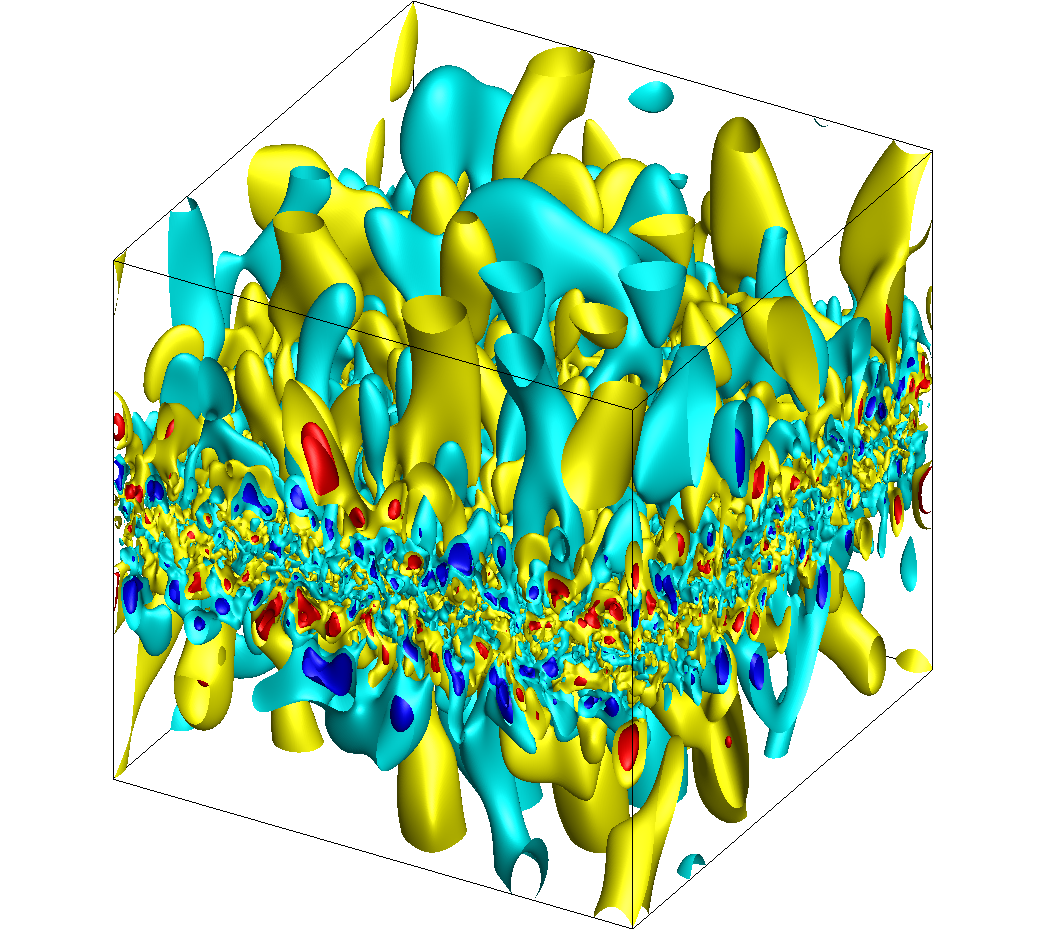}
	\end{minipage}
	\vfill
	\begin{minipage}[b]{1.0\textwidth}
		\centering
		\includegraphics[trim=2pt 2pt 2pt 2pt, clip, width=.3\textwidth]{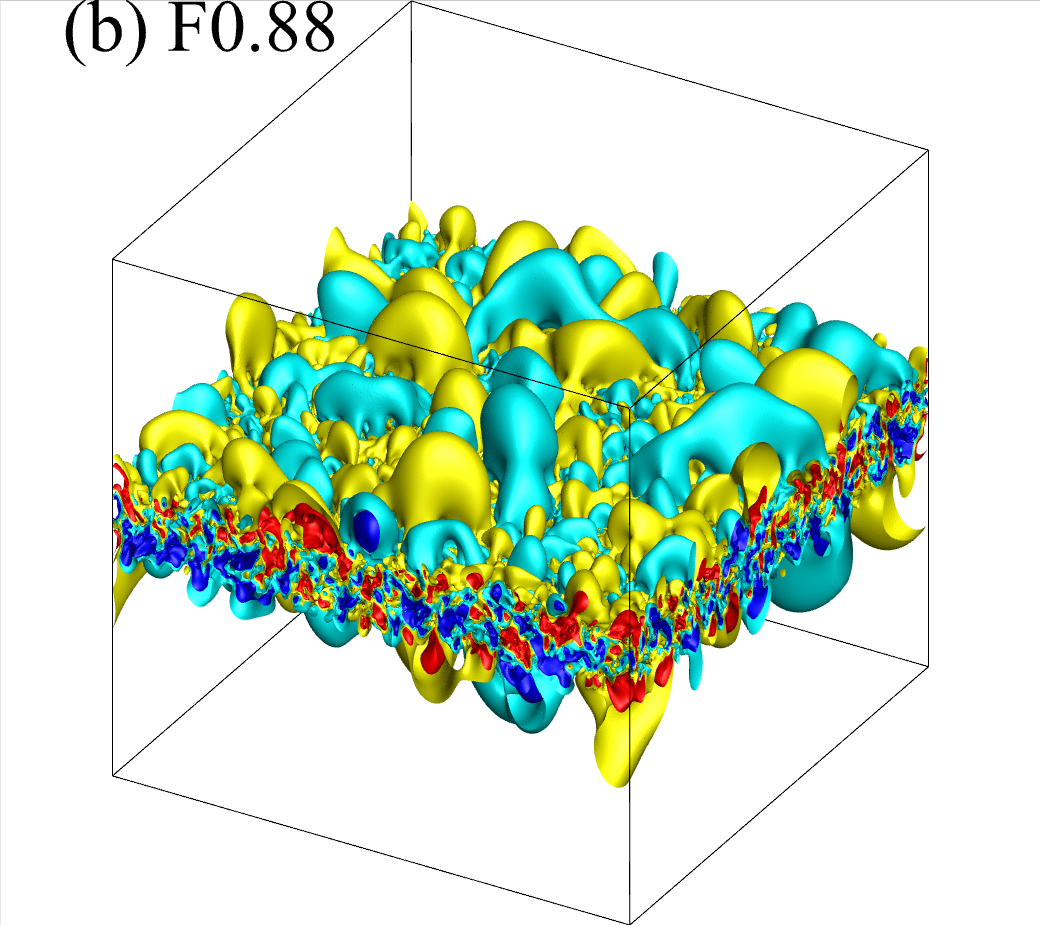}
		\includegraphics[trim=2pt 2pt 2pt 2pt, clip, width=.3\textwidth]{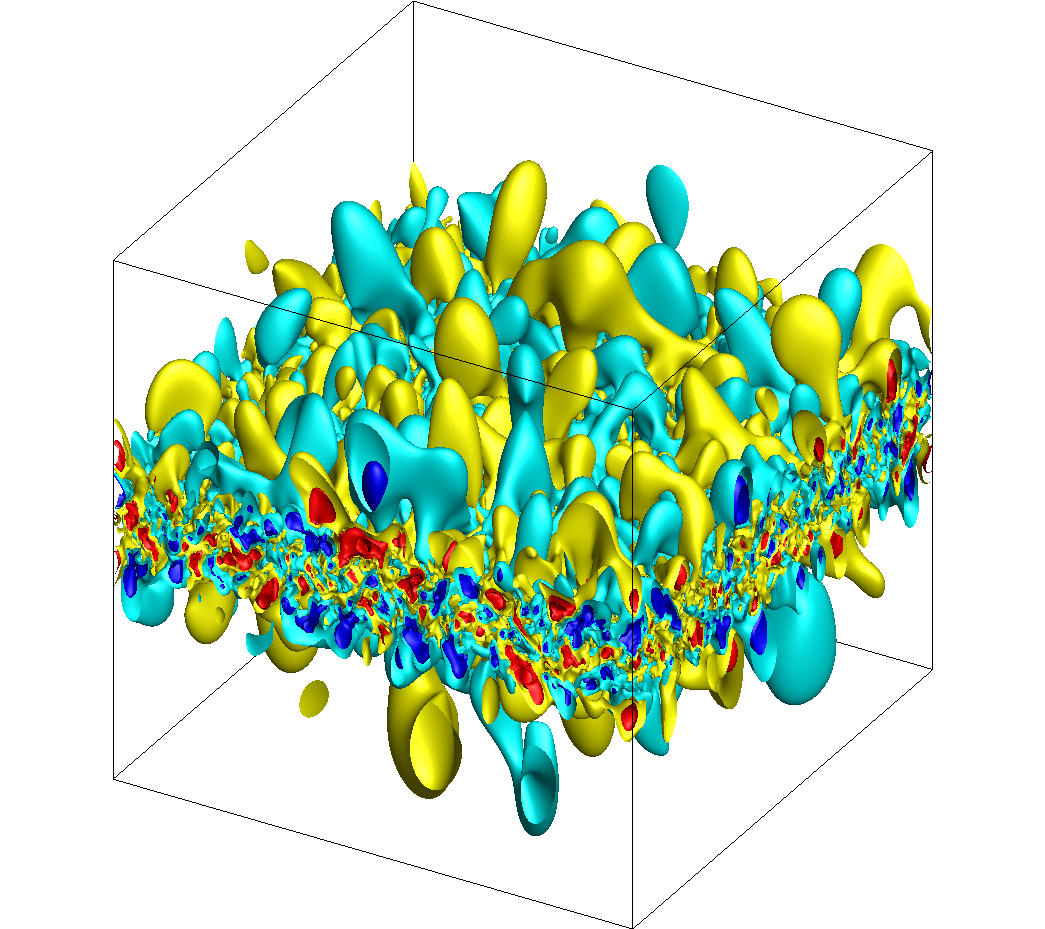}
		\includegraphics[trim=2pt 2pt 2pt 2pt, clip, width=.3\textwidth]{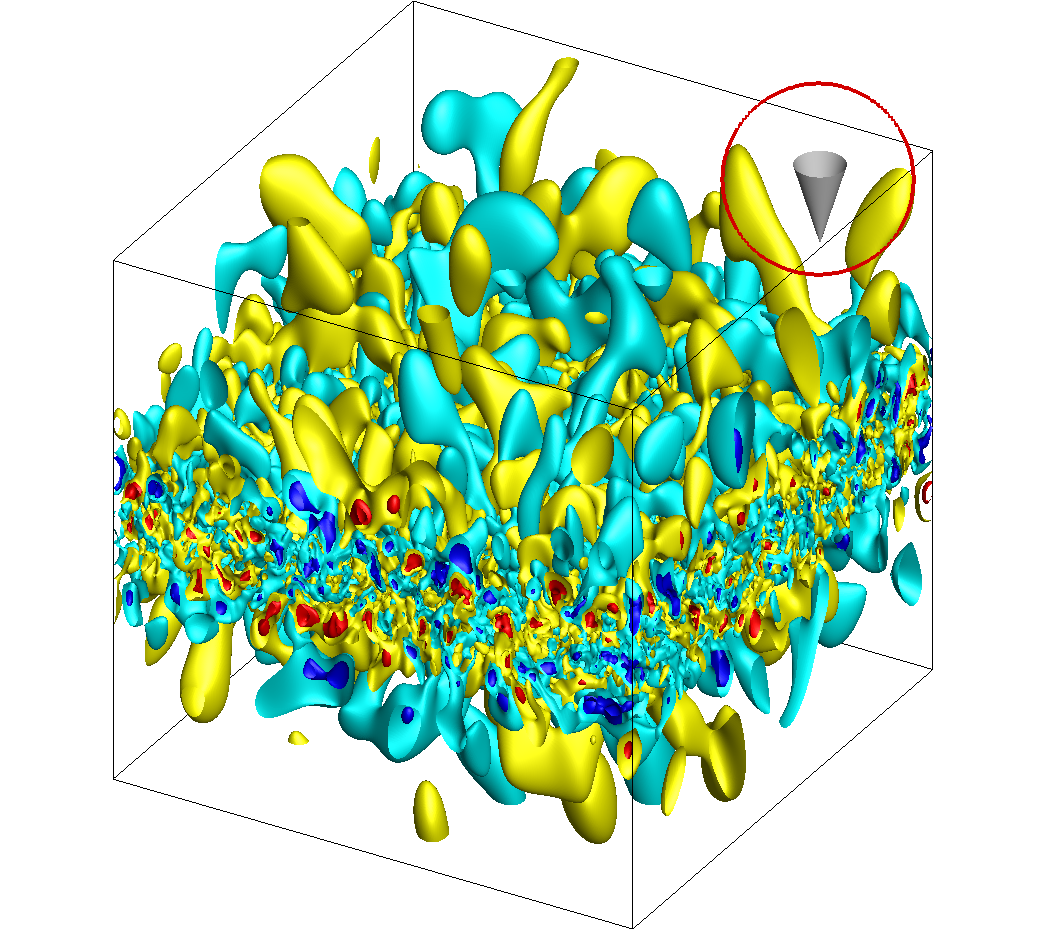}
	\end{minipage}
	\vfill
	\begin{minipage}[b]{1.0\textwidth}
		\centering
		\includegraphics[trim=2pt 2pt 2pt 2pt, clip, width=.3\textwidth]{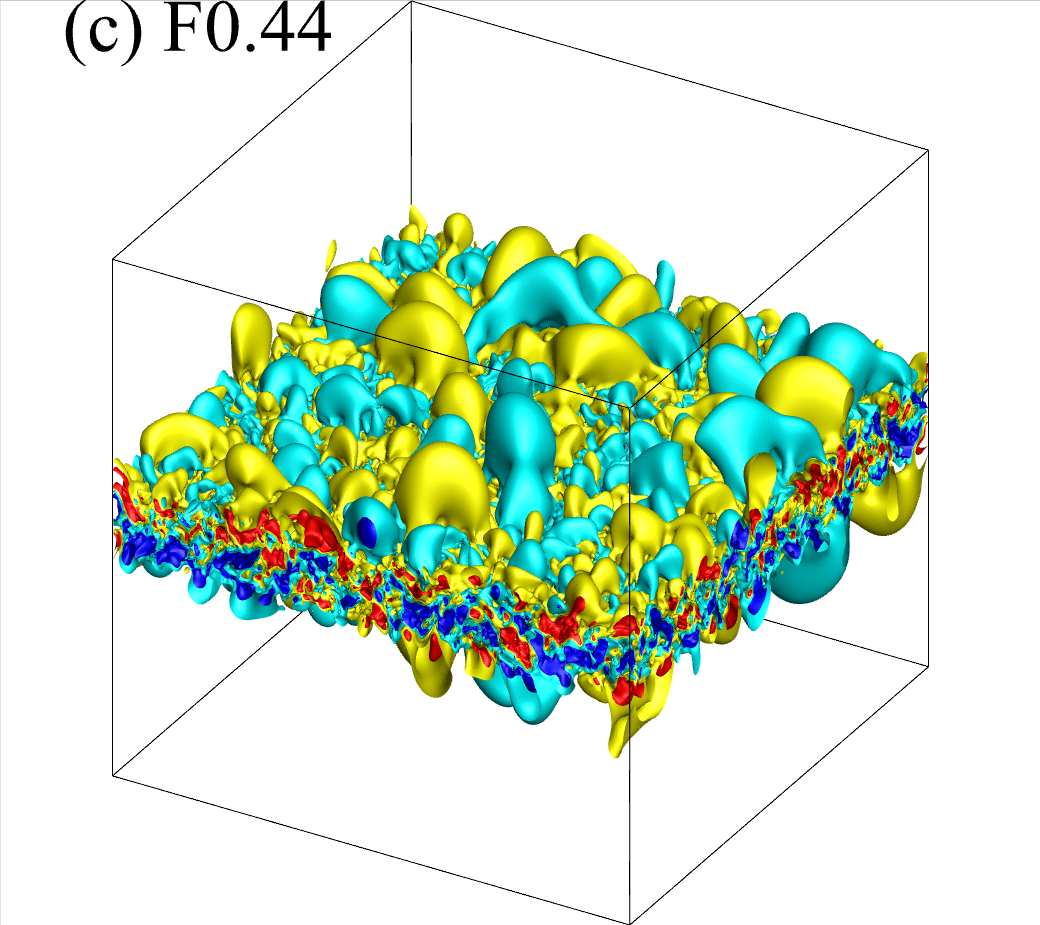}
		\includegraphics[trim=2pt 2pt 2pt 2pt, clip, width=.3\textwidth]{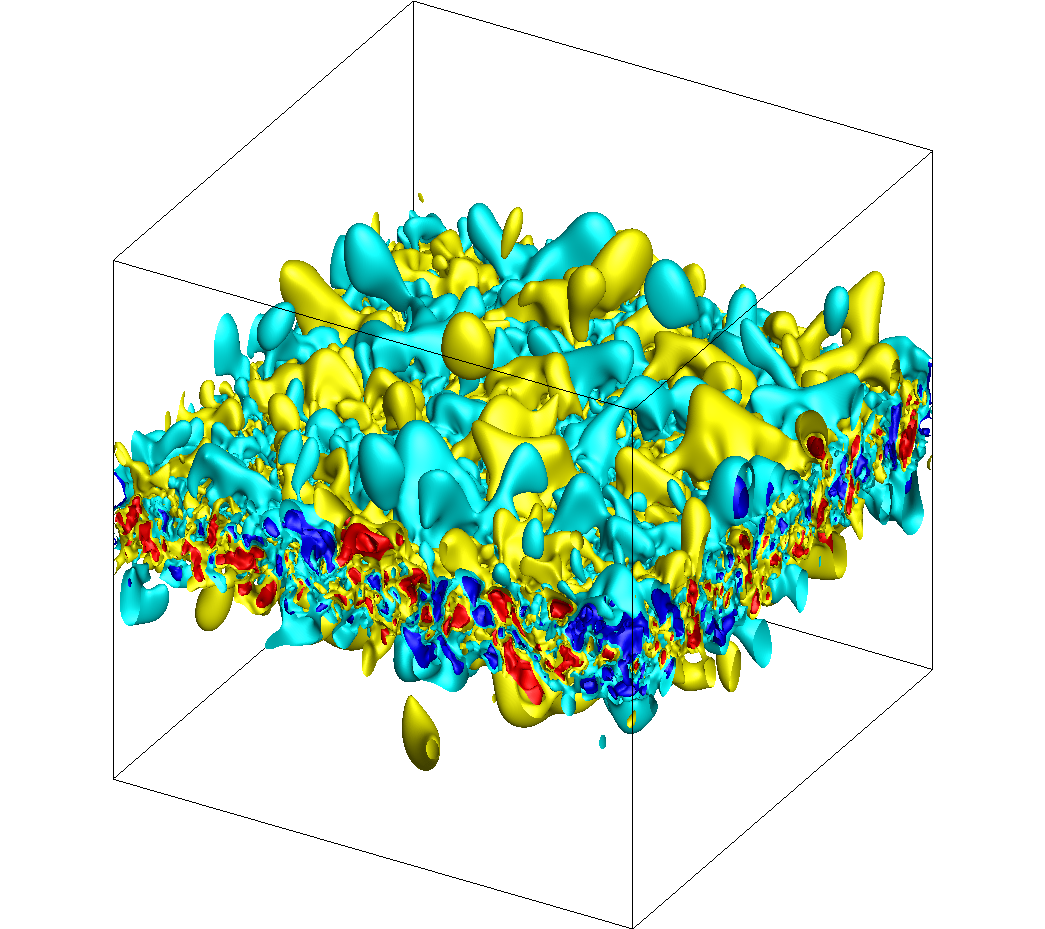}
		\includegraphics[trim=2pt 2pt 2pt 2pt, clip, width=.3\textwidth]{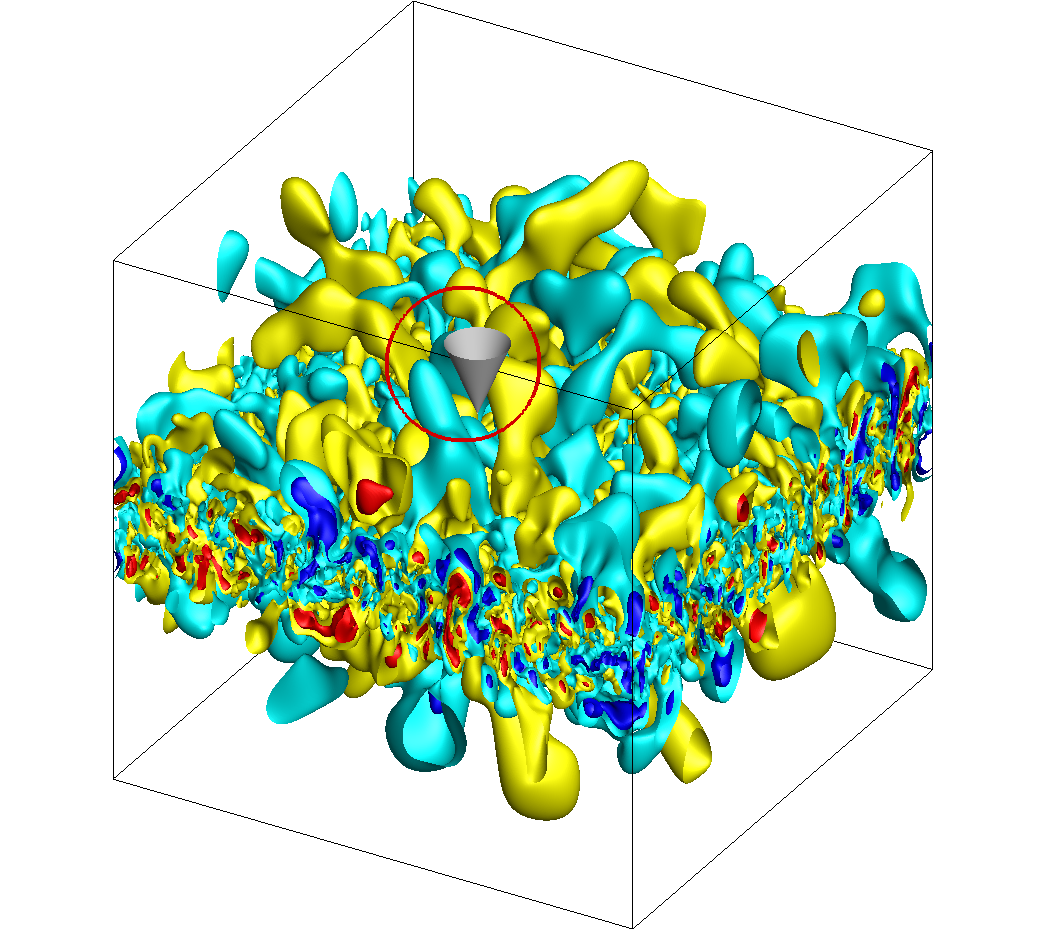}
	\end{minipage}
	\vfill
	\begin{minipage}[b]{1.0\textwidth}
		\centering
		\includegraphics[trim=2pt 2pt 2pt 2pt, clip, width=.3\textwidth]{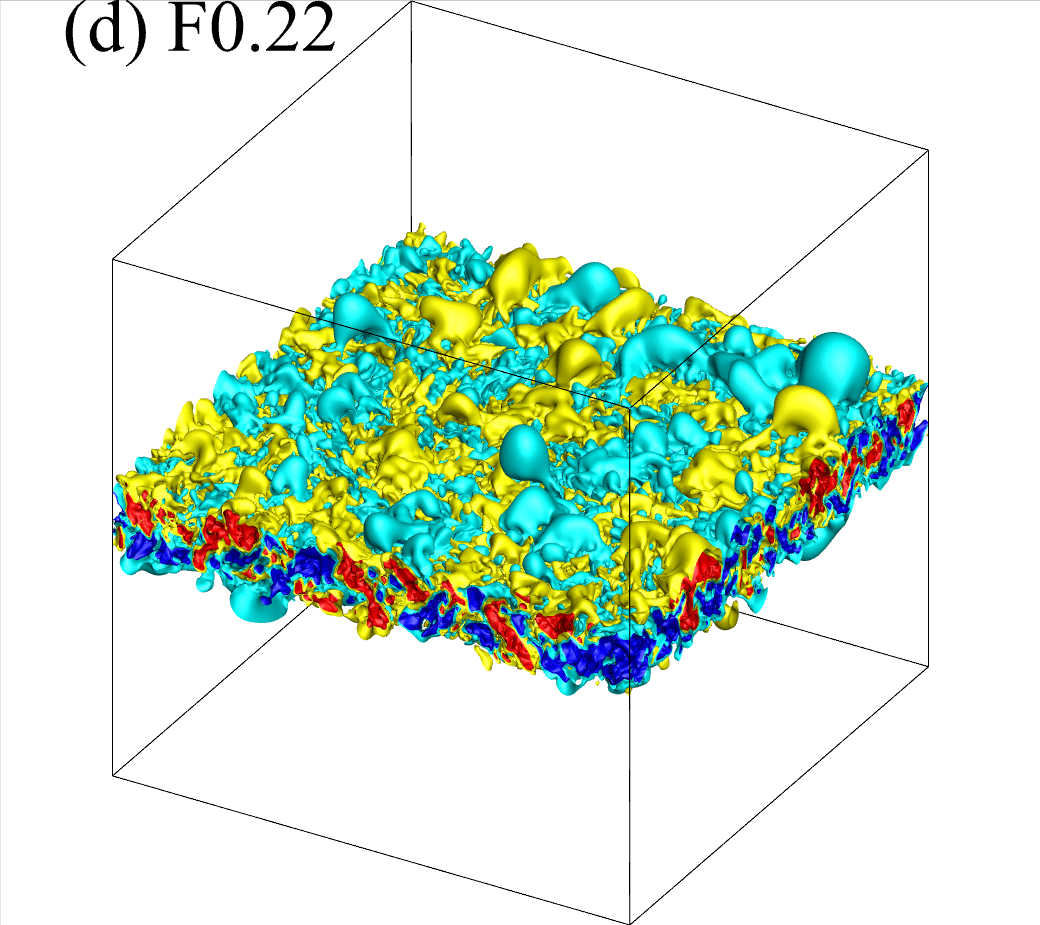}
		\includegraphics[trim=2pt 2pt 2pt 2pt, clip, width=.3\textwidth]{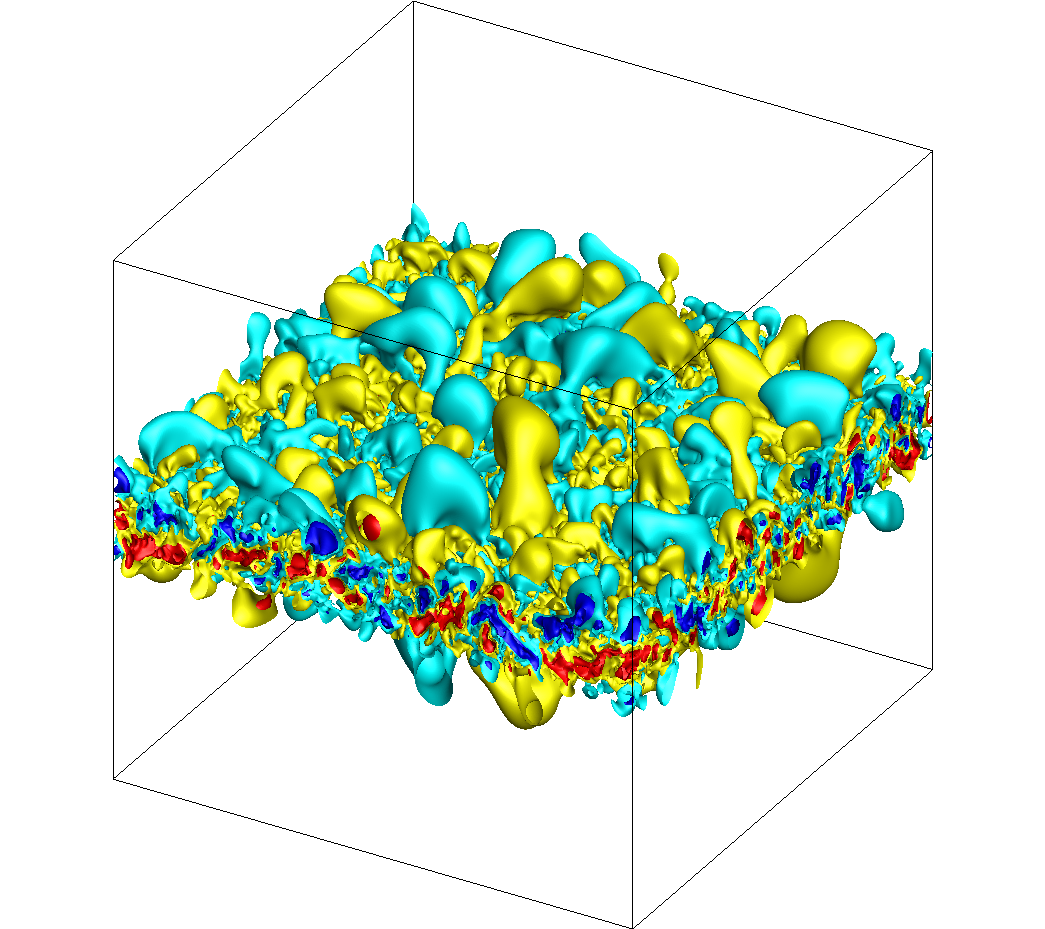}
		\includegraphics[trim=2pt 2pt 2pt 2pt, clip, width=.3\textwidth]{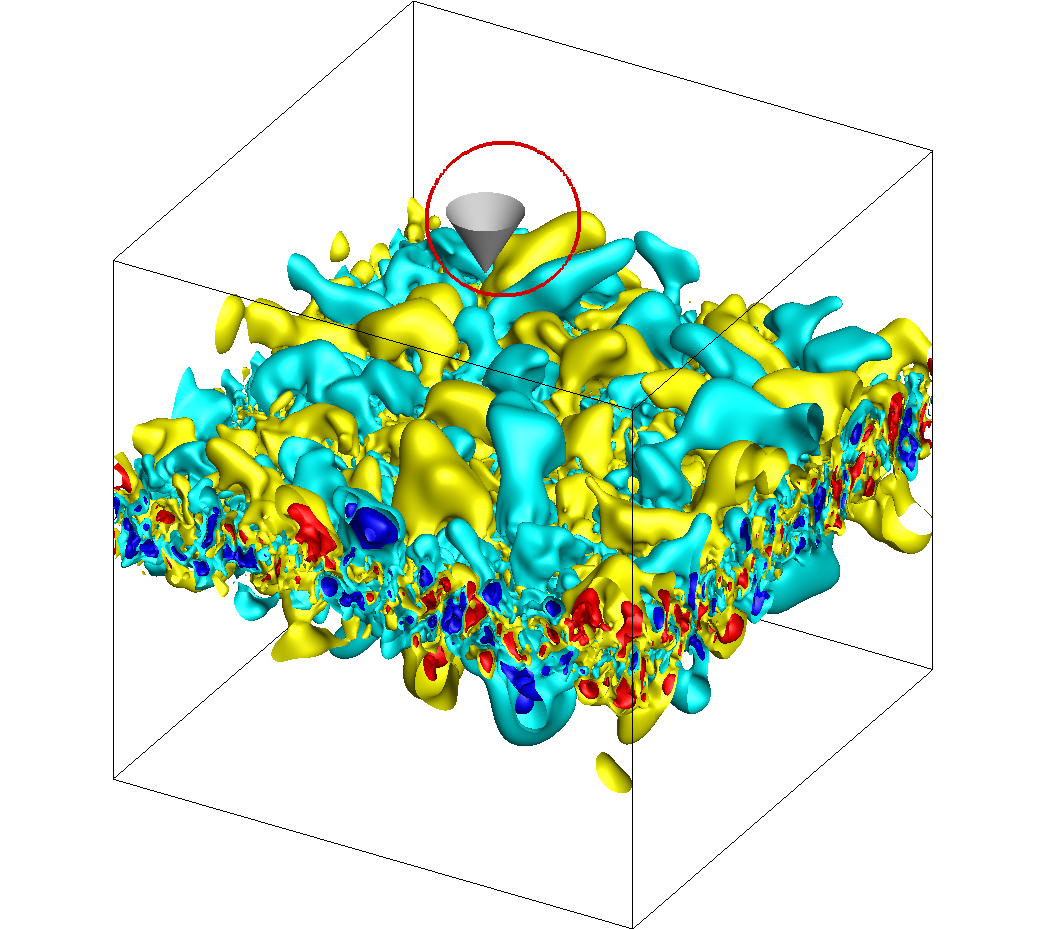}
	\end{minipage}
	\vfill
	\begin{minipage}[b]{1.0\textwidth}
		\centering
		\includegraphics[trim=2pt 2pt 2pt 2pt, clip, width=.3\textwidth]{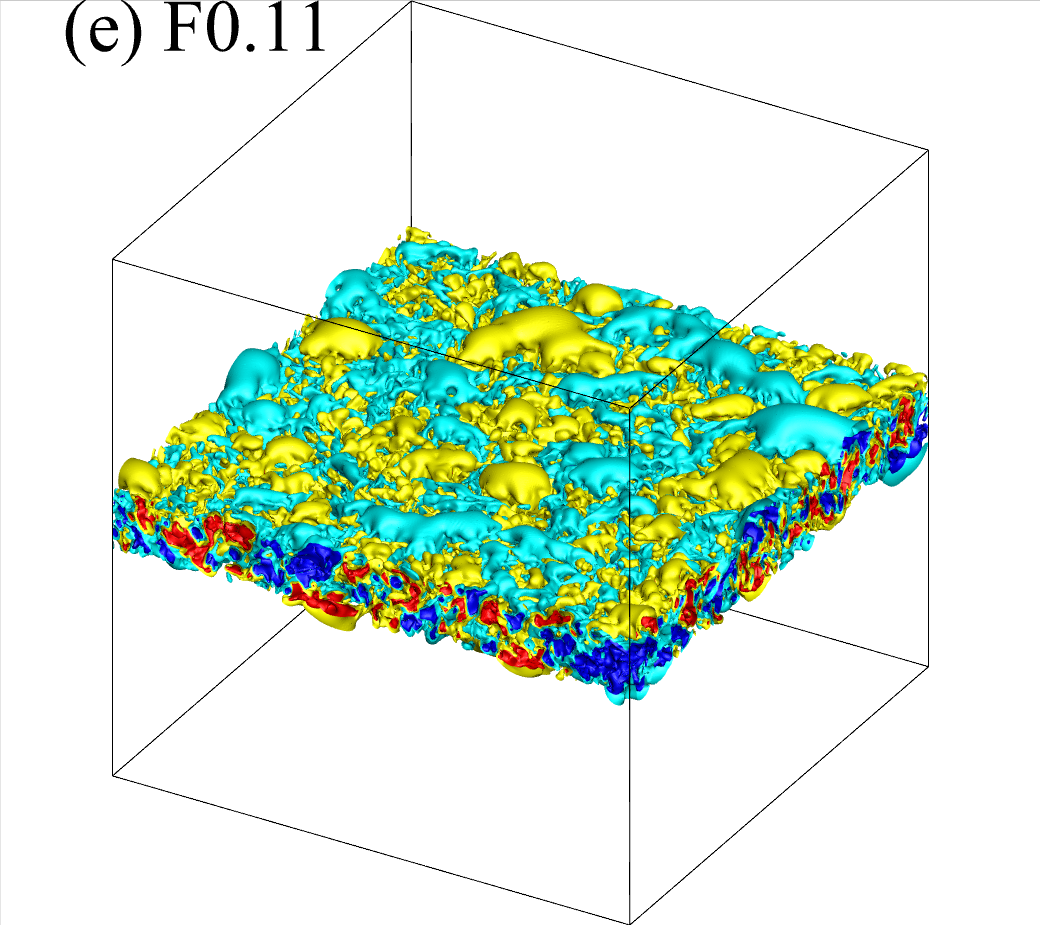}
		\includegraphics[trim=2pt 2pt 2pt 2pt, clip, width=.3\textwidth]{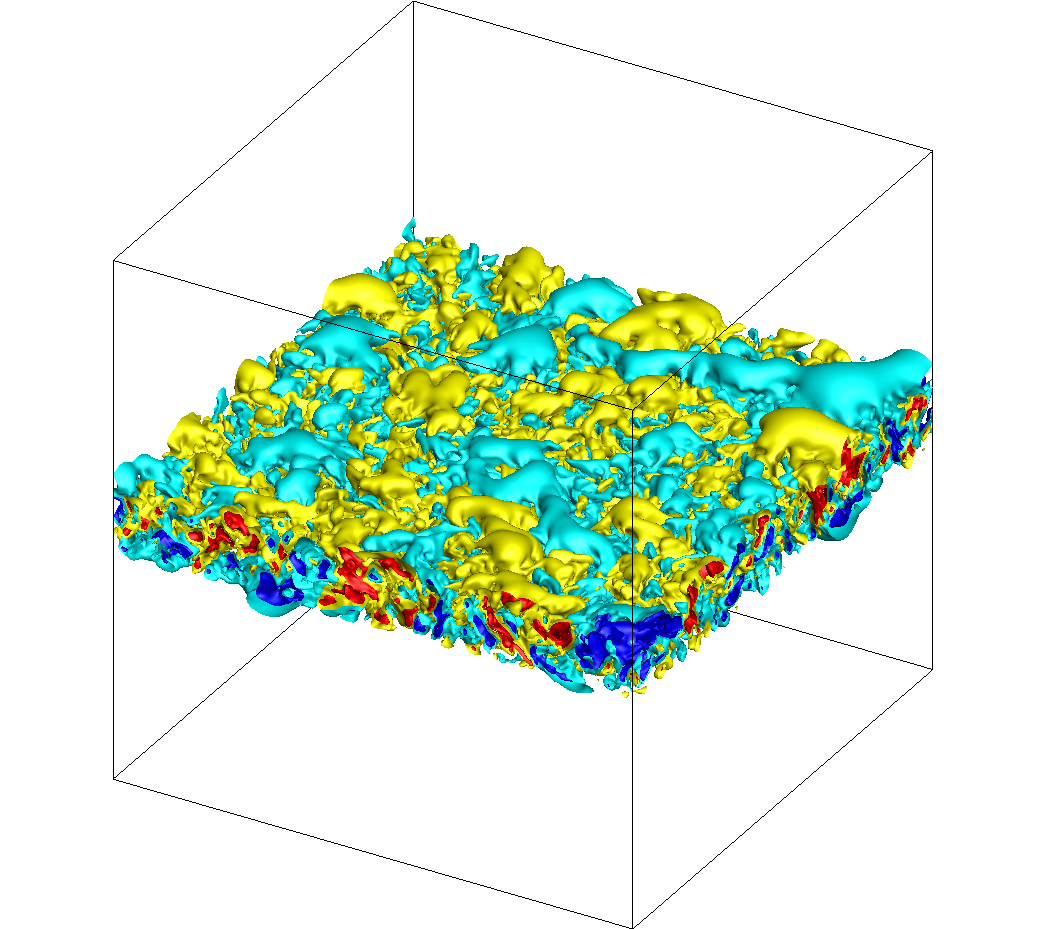}
		\includegraphics[trim=2pt 2pt 2pt 2pt, clip, width=.3\textwidth]{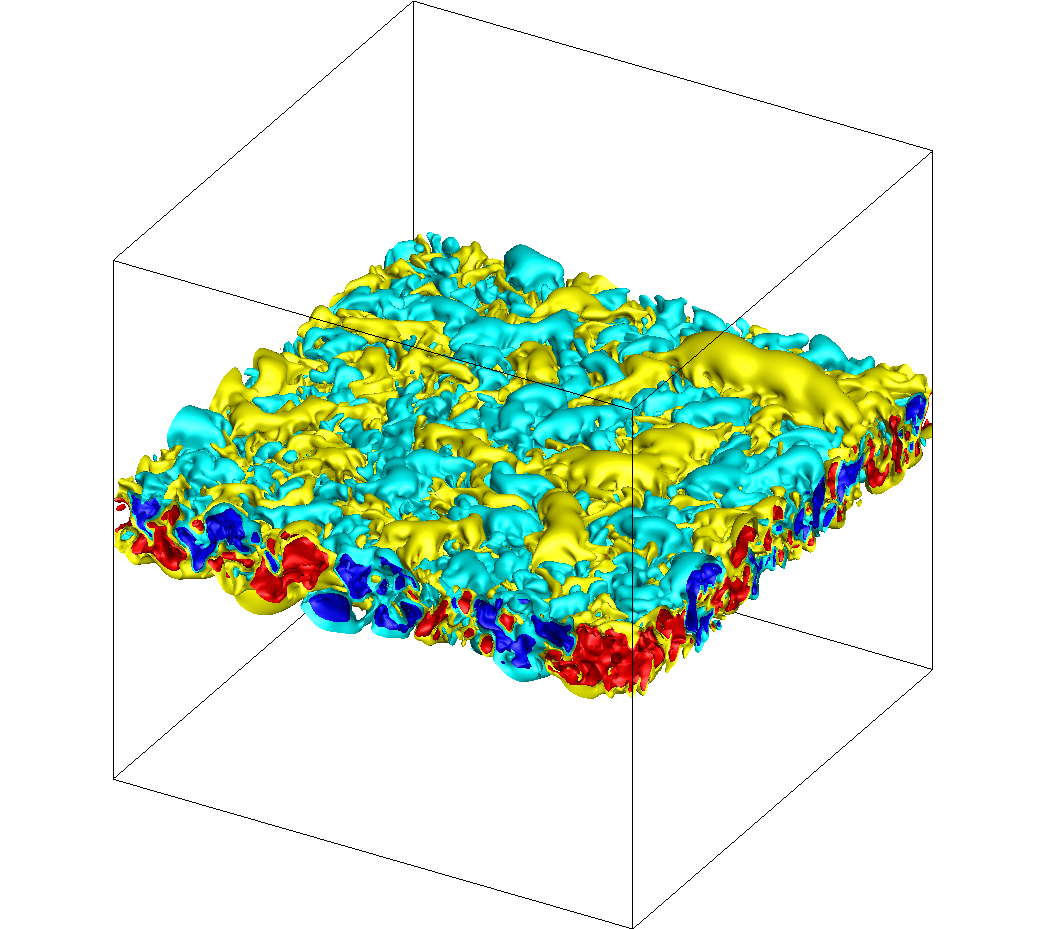}
	\end{minipage}
	\caption{\hl{The} evolution of the iso-surfaces of $u_z$ for the cases: (\textbf{a}) R0.11, (\textbf{b}) F0.88, (\textbf{c}) F0.44, (\textbf{d}) F0.22 and (\textbf{e}) F0.11 at $t/t_f=4.2$ (left), $8.4$ (middle) and $12.7$ (right). The color scheme is as follows: $0.24u_0$ in yellow, $0.50u_0$ in red, $-0.24u_0$ in cyan, and $-0.50u_0$ in blue. Cones \TL{(highlighted by red circles)} are plotted to indicate the predicted tilt angle of flow structures from the linear theory.}\label{fig:env_iso_uz}%MDPI: Please add explanation for the red circles in the figure caption % Reply: 8. Thanks. THe explanation has been added. Moreover, we adjusted the position (\centering) of the first 3 rows of the subfigures.
\end{figure}

To show the evolution of the turbulent cloud structures, Figure \ref{fig:env_iso_uz} shows iso-surfaces of $u_z$ for all simulations at different times. For the case R0.11 under pure rotation, \mbox{Figure \ref{fig:env_iso_uz}\TL{\textbf{a}}} shows the emergence of vertical columnar structures from the turbulent cloud, which extend into the quiescent region, in agreement with previous experimental results \citep{davidson2006evolution} and numerical study \citep{ranjan2014evolution}. For the weakly stratified case F0.88, Figure \ref{fig:env_iso_uz}\TL{\textbf{b}} shows subtle differences in the flow structures at $t/t_f = 8.4$ {(middle panels)} and more pronounced variations at $t/t_f = 12.7$ {(right panels)} compared to the case R0.11.

For the moderately stratified cases F0.44 and F0.22, Figure \ref{fig:env_iso_uz}\TL{\textbf{c},\textbf{d}} show that the columnar structures tend to deviate from a purely vertical orientation as they extend into the quiescent region. This tilt becomes more pronounced with increasing stratification. A comparison of Figure \ref{fig:env_iso_uz}\TL{\textbf{a}--\textbf{e}} shows that, at equivalent time intervals, the vertical growth of these structures decreases with decreasing $Fr$. This suggests that strong density stratification limits the vertical expansion of the cloud, consistent with the observation in experiments \citep{davies1991generation}.

In the case F0.11 (Figure \ref{fig:env_iso_uz}\TL{\textbf{e}}), the rotation frequency and the Brunt--V\"ais\"al\"a frequency are identical. As a result, no structures emerge from the turbulent cloud, suggesting minimal energy emission from the cloud. This behavior mirrors the case without both rotation and stratification, as documented in Ref. \citep{ranjan2014evolution}. A plausible interpretation is that the stratification strength is sufficient to completely inhibit the vertical expansion of the turbulent cloud in the presence of system rotation. It is noteworthy that Equation (\ref{equ:fig}) for this scenario yields $\varpi=2\Omega=N$, implying uniform frequencies of inertial-gravity waves in all directions. {This monochromatic behavior is consistent with systems in which both rotation and stratification are absent.}%Thus, in the absence of both rotation and stratification, the system exhibits isotropy.

\subsection{Are the Flow Structures Formed by Inertial-Gravity Waves?}\label{subsec:flow_structures_igw}

Columnar structures from a turbulent cloud under rotation have been identified as inertial waves \citep{ranjan2014evolution}, while the large-scale pancake structures from a stratified turbulent cloud are recognized as internal gravity waves \citep{maffioli2014evolution}. This section examines whether the flow structures discussed in Section \ref{subsec:evo_flow_struct} from the stratified turbulent cloud under rotation can be attributed to inertial-gravity waves.

\begin{figure}[H]
	\begin{minipage}[b]{.5\textwidth}
		\includegraphics[trim=2pt 2pt 2pt 2pt, clip, width=1.0\textwidth]{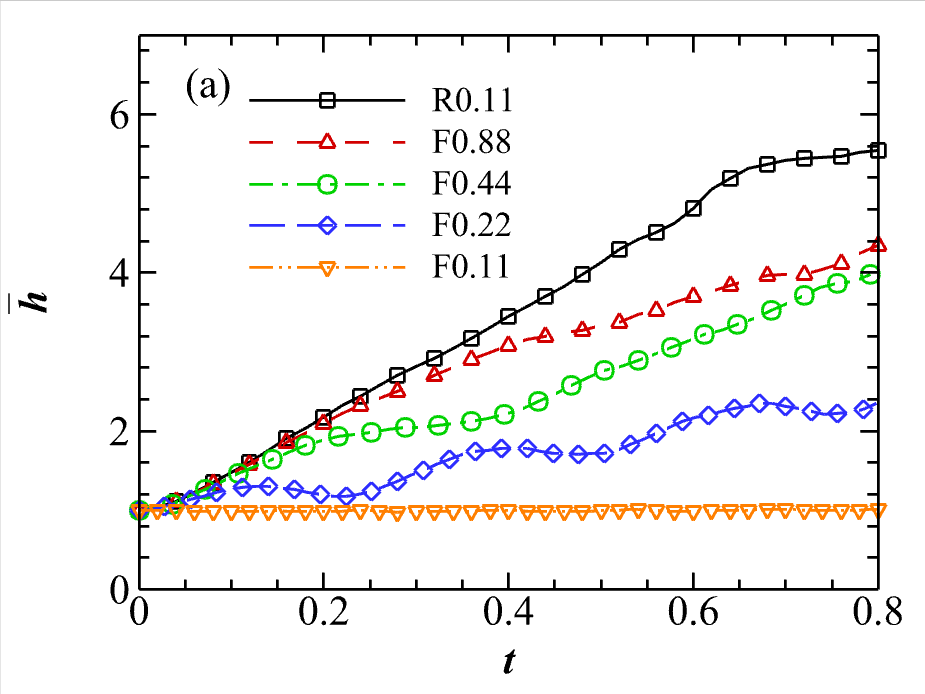}
	\end{minipage}
	\hfill
	\begin{minipage}[b]{.5\textwidth}
		\centering
		\includegraphics[trim=2pt 2pt 2pt 2pt, clip, width=1.0\textwidth]{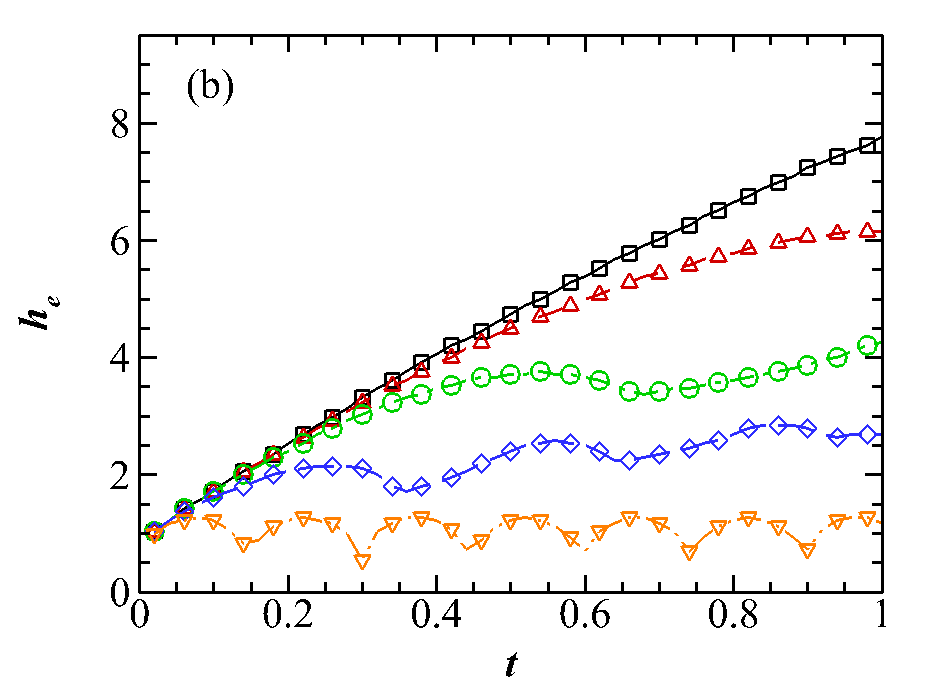}
	\end{minipage}
	\caption{(\textbf{a}) The mean wave-cloud thickness $\overline{h}$ of turbulent clouds in the DNS. (\textbf{b}) The vertical extent $h_e$ of a Gaussian eddy evolving under the same $Ro$ and $Fr$ as those in the DNS.}\label{fig:thickness}
\end{figure}

In all simulations, the growth of flow structures in different directions leads to the expansion of turbulent clouds in the vertical direction, which can be quantified by tracking the boundary of the $|u_z|$ iso-surfaces with time. The reason we choose $u_z$ to indicate structure is that it satisfies the wave equation as described by (\ref{equ:luz}). In particular, given a threshold $|u_z^b|$, the upper and lower boundaries of the cloud can be obtained from the positions where $|u_z|=|u_z^b|$. We can therefore calculate the vertical extent of the cloud, $h(x,y,t)$, for each point on the \emph{x}-\emph{y} plane and time $t$. The mean wave-cloud thickness $\overline{h}(t)$ is defined as the average of $h(x,y,t)$ over $x$ and $y$. Figure \ref{fig:thickness}\TL{\textbf{a}} shows the mean wave-cloud thickness for all the simulations where we choose $|u_z^b|=0.35u_0$. The results remain robust when the threshold varies around this value, and this threshold is used consistently throughout the analysis in this paper. For the purely rotating case R0.11, the mean wave-cloud thickness shows a linear growth with time, resulting from the inertial waves with $k_z=0$ traveling at the group velocity $c_g=2\Omega⁄k_h$ \citep{ranjan2014evolution}. For the case F0.88 with weak stratification, the mean wave-cloud thickness is close to that in the case R0.11 at $t\le0.2$, indicating that the stratification does not affect the cloud before $t=0.2$. This is because the linear time scale corresponding to the rotation, $t_f=1/(2\Omega)$, is smaller than that corresponding to the stratification, $t_N=1/N$. When $t>0.2$, the mean wave-cloud thickness shows a smaller linear growth rate. For the cases F0.44 and F0.22 with stronger stratification, the behavior of the mean wave-cloud thicknesses can be interpreted as the superposition of a linear growth and an oscillation with a fixed frequency over time. In addition, the frequency is larger in the case F0.22. The mean wave-cloud thickness in the case F0.11 is almost constant with time, which is consistent with the observation in Figure \ref{fig:env_iso_uz}\TL{\textbf{e}}.

\begin{figure}[H]
	\begin{minipage}[b]{.5\textwidth}
		\includegraphics[trim=2pt 2pt 2pt 2pt, clip, width=1.0\textwidth]{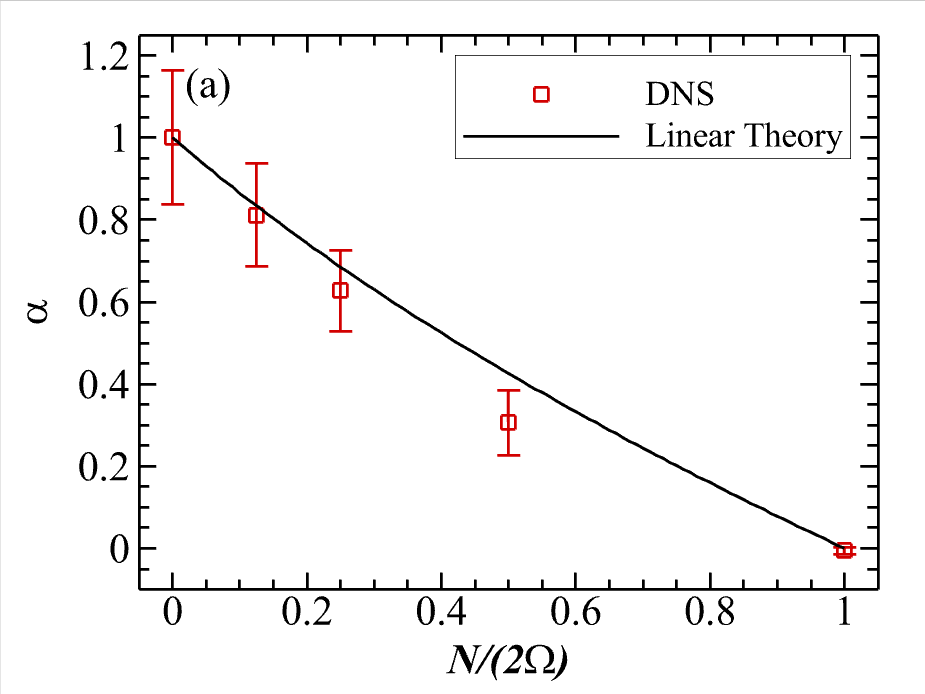}
	\end{minipage}
	\hfill
	\begin{minipage}[b]{.5\textwidth}
		\includegraphics[trim=2pt 2pt 2pt 2pt, clip, width=1.0\textwidth]{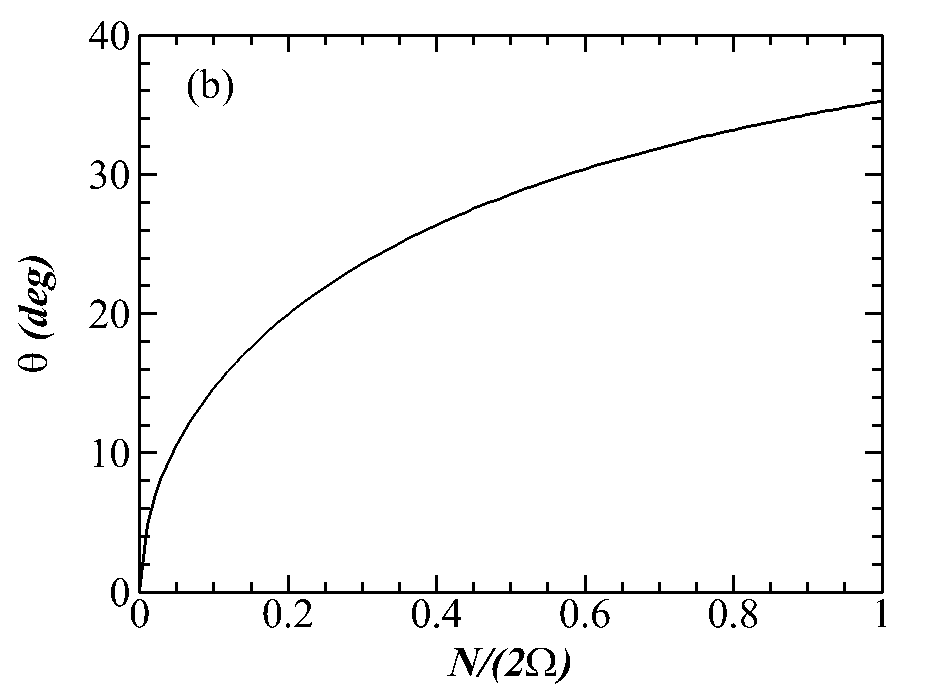}
	\end{minipage}
	\caption{(\textbf{a}) Normalized vertical growth rates of the turbulent cloud, $\alpha$, obtained from the DNS results and the linear theory. (\textbf{b}) The angle $\theta$ between the flow structure and the vertical direction as predicted by the linear theory.}\label{fig:LinearTheory}
\end{figure}

To investigate whether the flow structures are closely related to the inertial-gravity waves, we compare the results from the DNS with those predicted by the linear theory. Since the initial turbulent cloud can be viewed as a distribution of random vortex blobs of different shapes and sizes, we consider the linear evolution of a Gaussian vortex satisfying Equation (\ref{equ:Gev}) under the same Rossby and Froude numbers. We set $\Lambda=2.31$, $\delta=0.125$, and utilize $|u_z^b|=0.002\Lambda\delta$ to determine the vertical extent of the eddy, $h_{e}$. This extent is defined as the largest vertical distance between the positions where $|u_z|=|u_z^b|$. Figure \ref{fig:thickness}\TL{\textbf{b}} shows that the results for $h_{e}$ align qualitatively with those of $\overline{h}$ in Figure \ref{fig:thickness}\TL{\textbf{a}}, suggesting that the expansion of the turbulent cloud may be prominently influenced by the linear mechanism. {Note that the difference for the case F0.11, where the curve appears flat in Figure \ref{fig:thickness}\TL{\textbf{a}} and wave-like in Figure \ref{fig:thickness}\TL{\textbf{b}}, arises because $\overline{h}$ represents an average over various $x$ and $y$, whereas $h_{e}$ illustrates the results for a single eddy.}
We now show quantitatively that the flow structures are closely related to inertial-gravity waves. For a vortex blob in the initial turbulent cloud, {we suppose that} it is characterized by a length scale of $1/k$. Then its rescaled group velocity in the vertical direction is
\begin{linenomath}
    \begin{equation}
        \tilde{c}_{g,z}=kc_{g,z}=\pm\frac{2\Omega(1-(N/(2\Omega))^2)(k_z/k_h)}{\sqrt{(N/(2\Omega))^2+(k_z/k_h)^2}(1+(k_z/k_h)^2)}.
    \end{equation}
\end{linenomath}
Given the ratio of the Brunt--V\"ais\"al\"a frequency to the rotation frequency $N/(2\Omega)$, it is determined that when
\begin{linenomath}
    \begin{equation}
        \left(\frac{k_z}{k_h}\right)^2=\frac{1}{4}\left[-(N/(2\Omega))^2+\sqrt{(N/(2\Omega))^4+8(N/(2\Omega))^2}\right],
    \end{equation}
\end{linenomath}
$\tilde{c}_{g,z}$ reaches the maximum $\tilde{c}_{g,z}^{max}$, which is the vertical growth rate of the turbulent cloud predicted by the linear inertial-gravity waves. Furthermore, the corresponding value
of $k_z/k_h=(k_z/k_h)_0$ characterizes the tilt angle of the flow structure. The vertical growth rate of the turbulent cloud in the DNS can be determined from the growth rate of the mean wave-cloud thickness, $\overline{h}(t)$, as shown in Figure \ref{fig:thickness}\TL{\textbf{a}}. The rate is calculated using a linear fitting method. In Figure \ref{fig:LinearTheory}\TL{\textbf{a}} we present two vertical growth rates: $\alpha_{DNS}$ derived from DNS and $\alpha_{LT}$ derived from linear theory. To facilitate comparison, both rates are normalized by their values at $N/2\Omega=0$. Figure \ref{fig:LinearTheory}\TL{\textbf{b}} shows the angle between the flow structure and the vertical direction, $\theta=\arctan[(k_z/k_h)_0]$, predicted by linear theory for different $N/(2\Omega)$. The results are consistent with Figure \ref{fig:env_iso_uz}, where the angle becomes larger as $N/(2\Omega)$ increases. For quantitative comparison, we also plot cones in the third columns of Figure \ref{fig:env_iso_uz}\TL{\textbf{b}--\textbf{d}}, where the angle between the cone surface and the vertical axis is equal to $\theta$. It is notable that there is a strong agreement between the DNS results and the predictions of linear theory. However, for the case F0.22, both $\alpha$ and $\theta$ are slightly underestimated by the linear theory. This discrepancy could be due to the interactions between waves and turbulence, especially given the small vertical expansion of the cloud in this case. It is important to note that in scenarios where both rotation and stratification are present, the frequency of the inertial-gravity waves (\ref{equ:fig}) tied to $\tilde{c}_{g,z}^{max}$ is not zero. This explains the oscillatory component observed in $\overline{h}$ for the cases F0.44 and F0.22. When only rotation or stratification is present, the maximum group velocity always coincides with inertial (or gravity) waves of zero frequency. Under these circumstances, tools such as two-dimensional energy spectrum analysis can be used effectively, given the consistent phase of $u_z$ \citep{ranjan2014evolution} or $u_x$ \citep{maffioli2014evolution} at the edge of the turbulent cloud.

\subsection{Wave-Dominated and Turbulence-Dominated Regions}\label{subsec:particle_tracking}

In Section \ref{subsec:flow_structures_igw} it is highlighted that the flow structures emerging from the turbulent cloud show a strong association with inertial-gravity waves. In stratified turbulence, PV has been used to identify turbulent/non-turbulent interfaces \citep{maffioli2014evolution, watanabe2016turbulent}. Our study, however, uses Lagrangian particle tracking to measure the vertical extent of turbulent \mbox{advection \citep{ranjan2014evolution}}. Lagrangian particles act as passive tracers, with their position $\boldsymbol{r}$ at time $t$ given by \mbox{$\mathrm{d}\boldsymbol{r}/\mathrm{d}t=\boldsymbol{u}(\boldsymbol{r},t)$}. In this relationship, $\boldsymbol{u}$ represents the velocity of the fluid. To determine $\boldsymbol{u}$ at the particle positions, we use a 6th-order Lagrange interpolation. Furthermore, the positions of these particles are integrated using a 2nd-order Adams--Bashforth scheme.

\begin{figure}[H]
	\begin{minipage}[b]{.5\textwidth}
		\includegraphics[trim=2pt 2pt 2pt 2pt, clip, width=1.0\textwidth]{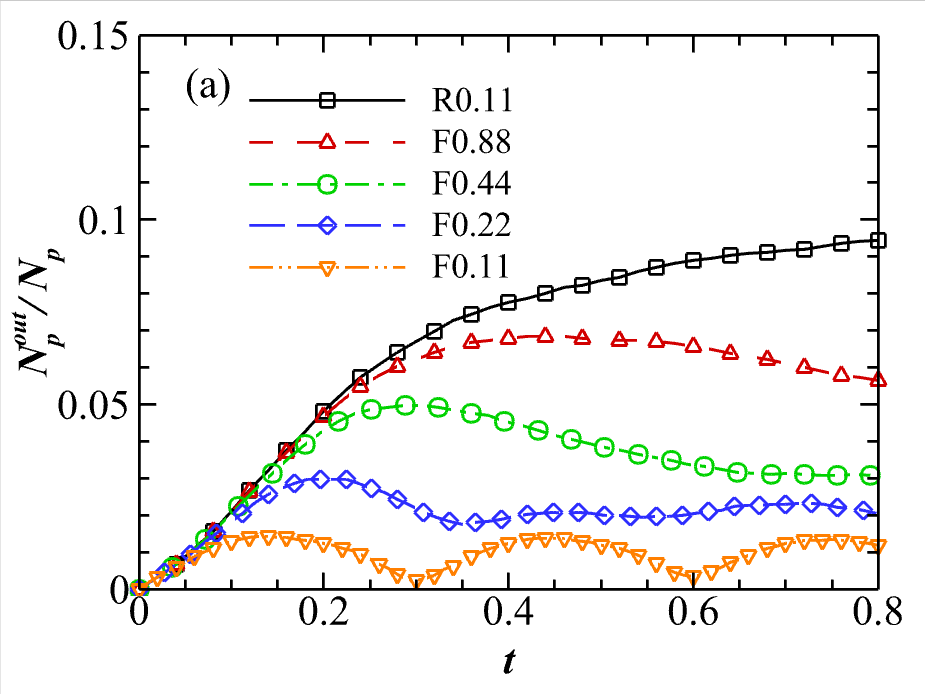}
	\end{minipage}
	\hfill
	\begin{minipage}[b]{.5\textwidth}
		\includegraphics[trim=2pt 2pt 2pt 2pt, clip, width=1.0\textwidth]{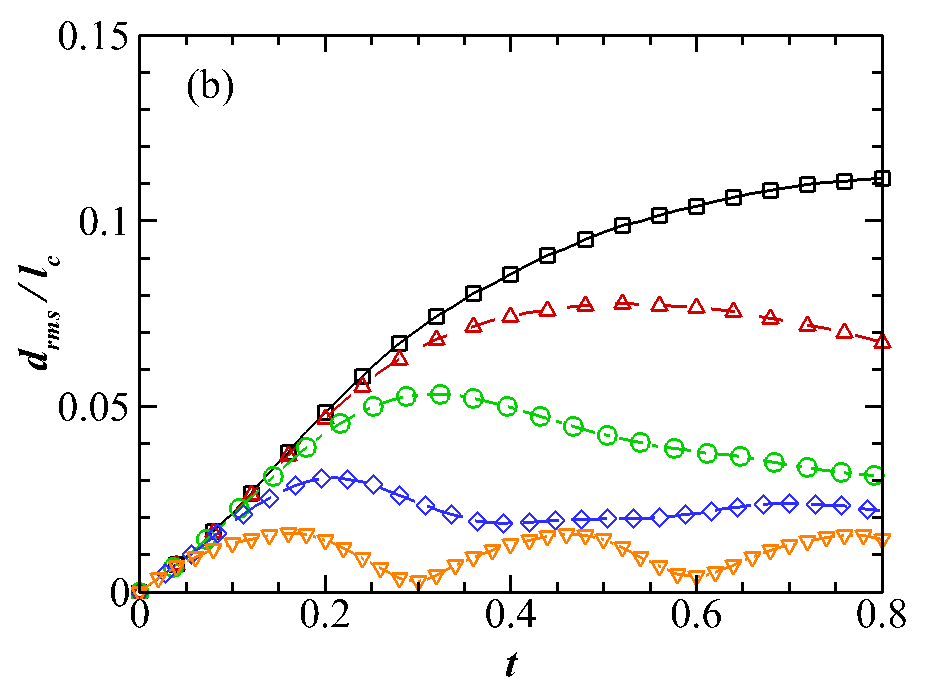}
	\end{minipage}
	\caption{(\textbf{a}) The ratio of particles outside the initial turbulent cloud. (\textbf{b}) The r.m.s. distance of particles outside the initial turbulent cloud to its boundary.}\label{fig:particle_tracking}
\end{figure}

Tracer particles with a number $N_p=65536$ are initialized randomly from a uniform distribution throughout the initial turbulent cloud, the vertical range of which is denoted as $[z_0,z_1]$. Let $N_p^{out}$ be the number of particles outside of the range of the initial turbulent cloud. The r.m.s. distance of these particles to the initial turbulent cloud can be calculated as
\begin{linenomath}
    \begin{equation}
        d_{rms}=\sqrt{\frac{1}{N_p^{out}}\sum_{p=1}^{N_p^{out}}d_p^2},
    \end{equation}
\end{linenomath}
where
\begin{linenomath}
    \begin{equation}\label{equ:dis_pc}
        d_p = \left\{
        \begin{array}{ll}
            z_p-z_1, & z_p>z_1\\[2pt]
            z_0-z_p, & z_p<z_0
        \end{array} \right.
    \end{equation}
\end{linenomath}
represents the distance of the particle $p$ to the initial turbulent cloud. We consider $[z_0-d_{rms}, z_1+d_{rms}]$ as the vertical range of the turbulence-dominated region and the other domain as the wave-dominated region. Figure \ref{fig:particle_tracking} shows the ratio of particles outside the initial turbulent cloud, $N_p^{out}/N_p$, and $d_{rms}$ normalized with the initial cloud thickness $l_c$. For the case R0.11, both $N_p^{out}/N_p$ and $d_{rms}/l_c$ increase with time, while for the cases F0.88 and F0.44 both quantities first increase and then decrease after reaching the maximum values, {likely approaching a stationary state due to the stable stratification.} However, in the cases F0.22 and F0.11, where the stratification is strong, both quantities immediately reach stationary states around small values.
Table \ref{tab:mnd} presents the maximum values of $N_p^{out}/N_p$ and $d_{rms}/l_c$ for all cases, where $\max(d_{rms}/l_c)$ can also be interpreted as the maximum vertical expansion of the initial turbulent cloud. It is observed that both quantities become smaller with increasing stratification, indicating the limitation of vertical turbulent advection by stratification.
\begin{table}[H]
	\begin{center}
	\caption{Maximum proportion of particles located outside the initial turbulent cloud, and the maximum vertical cloud expansion for the different cases.}
\setlength{\cellWidtha}{\textwidth/6-2\tabcolsep+0.4in}
\setlength{\cellWidthb}{\textwidth/6-2\tabcolsep-0.1in}
\setlength{\cellWidthc}{\textwidth/6-2\tabcolsep-0.1in}
\setlength{\cellWidthd}{\textwidth/6-2\tabcolsep-0.1in}
\setlength{\cellWidthe}{\textwidth/6-2\tabcolsep-0.1in}
\setlength{\cellWidthf}{\textwidth/6-2\tabcolsep+0.1in}
\scalebox{1}[1]{\begin{tabularx}{\textwidth}{>{\centering\arraybackslash}m{\cellWidtha}>{\centering\arraybackslash}m{\cellWidthb}>{\centering\arraybackslash}m{\cellWidthc}>{\centering\arraybackslash}m{\cellWidthd}>{\centering\arraybackslash}m{\cellWidthe}>{\centering\arraybackslash}m{\cellWidthf}}

            \toprule
			\textbf{Case} 	  & \textbf{R0.11}  & \textbf{F0.88}  & \textbf{F0.44}  & \textbf{F0.22 } & \textbf{F0.11} \\
            \midrule
			$\max(N_p^{out}/N_p)$ & $9.45\%$ & $6.86\%$ & $4.98\%$ & $2.99\%$ & $1.44\%$ \\
			$\max(d_{rms}/l_c)$ & $11.16\%$    & $7.77\%$    & $5.33\%$    & $3.08\%$    & $1.57\%$   \\
            \bottomrule
		\end{tabularx}}
		
		\label{tab:mnd}
	\end{center}
\end{table}

\begin{figure}[H]
	\begin{minipage}[b]{.5\textwidth}
		\includegraphics[trim=2pt 2pt 2pt 2pt, clip, width=1.0\textwidth]{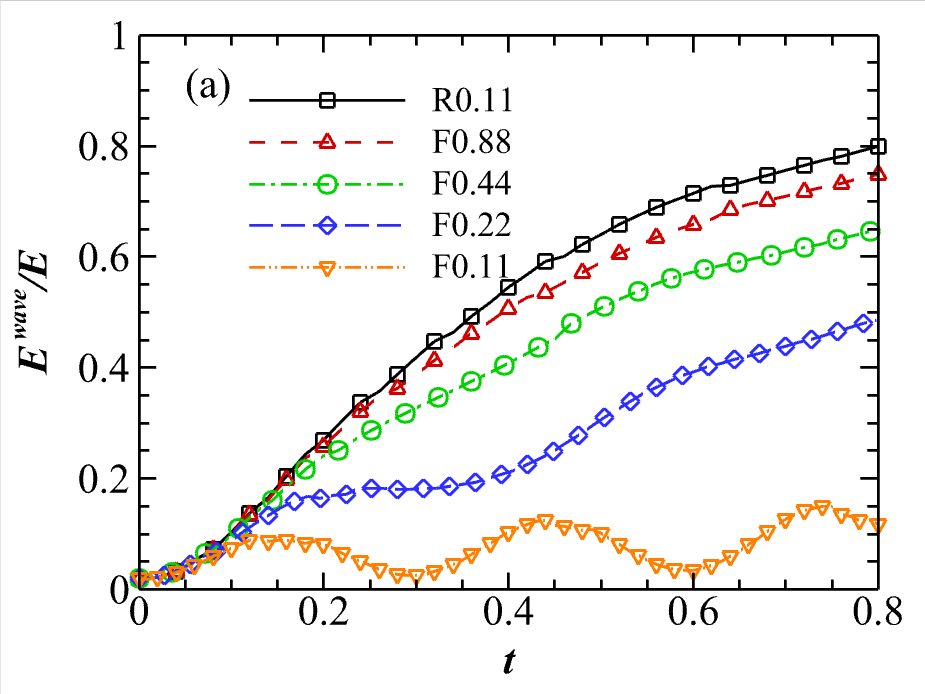}
	\end{minipage}
	\hfill
	\begin{minipage}[b]{.5\textwidth}
		\includegraphics[trim=2pt 2pt 2pt 2pt, clip, width=1.0\textwidth]{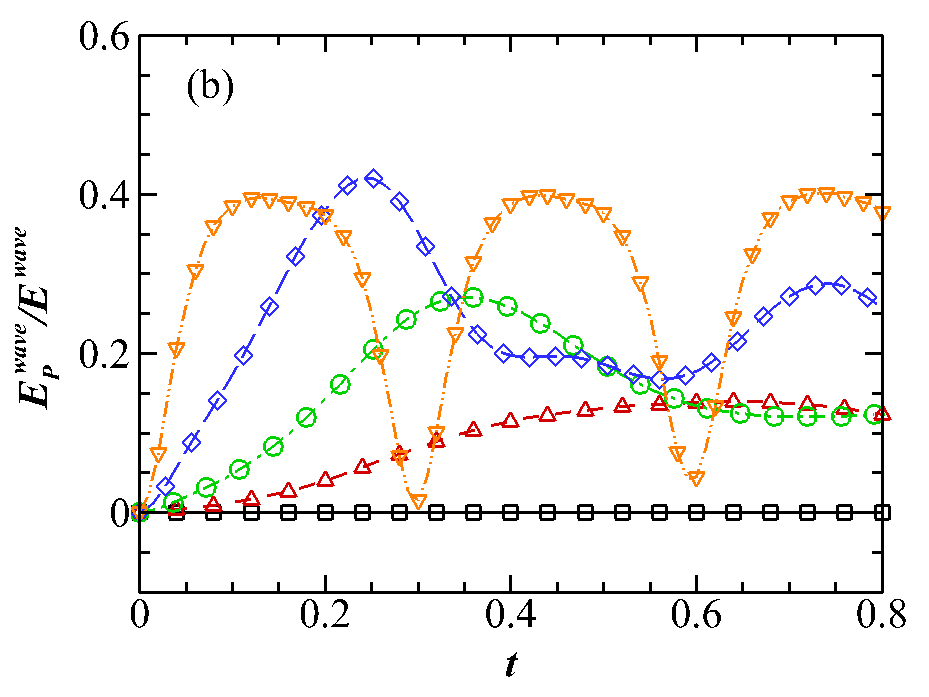}
	\end{minipage}
	\caption{(\textbf{a}) The energy ratio within the inertial-gravity waves. (\textbf{b}) The ratio of potential energy to total energy in the wave-dominated region.}\label{fig:wave_ener}
\end{figure}

Once the regions dominated by waves and turbulence are identified, we can determine the energy of the turbulent cloud carried away by the inertial-gravity waves, denoted as $E^{wave}$. This energy is equal to the energy within the wave-dominated region. Figure \ref{fig:wave_ener}\TL{\textbf{a}} shows that for cases where $Ro<Fr$, the ratio $E^{wave}/E$ exhibits a monotonic increase with time. In particular, for a given time point, a decrease in $Fr$ results in smaller values of $E^{wave}/E$. This suggests that under stronger stratification, inertial-gravity waves contain a smaller fraction of the energy. In contrast, for the scenario where $Ro=Fr$, the ratio $E^{wave}/E$ oscillates with a fixed pattern. Its values fluctuate between nearly zero and a peak of about $15\%$, indicating that the inertial-gravity waves only sporadically extract a small fraction of energy from the turbulent cloud. Thus, while inertial waves account for a significant fraction of the energy in a system dominated by rotation alone, this energy contribution diminishes with the introduction of stratification. Figure \ref{fig:wave_ener}\TL{\textbf{b}} shows the ratio of potential energy to total energy in the wave-dominated region, $E_P^{wave}/E^{wave}$. As stratification increases, potential energy becomes increasingly dominant and varies more frequently with time. This highlights the increasing importance of the kinetic-potential energy exchange and suggests that the dominant inertial-gravity waves exhibit greater phase velocities under increased stratification. It is worth noting that a similar kinetic-potential energy exchange phenomenon has been reported in {fully developed} rotating stratified turbulence \citep{li2020spectral}.

\section{Discussion and Conclusions}

In this study, we investigate the evolution of a stratified turbulent cloud influenced by rotation. By analyzing the behavior of a single eddy under both rotation and stratification, we observe clear deviations from the purely rotating scenario. In particular, within the stratified environment, we detect a spatial manifestation of a known PV mode, characterized as a stable vortex at the origin. Furthermore, the system transitions from exhibiting inertial waves to producing {inertial-gravity} waves, whose propagation relations depend on both rotation and stratification.

Numerical simulations have been performed to investigate the nonlinear evolution of the turbulent cloud influenced by different degrees of stratification under rotation. In cases where $Ro<Fr$, columnar structures emerge spontaneously from the turbulent cloud. These structures are vertically oriented in the case of pure rotation, but deviate from the vertical as stratification is introduced. As $Fr$ decreases, the tilt angle increases and the rate of cloud expansion decreases. Conversely, in the $Ro=Fr$ system, the turbulent cloud remains devoid of distinct structures, reminiscent of scenarios without rotation and stratification. The flow structures are observed to have characteristics consistent with inertial-gravity waves. This observation is supported by the agreement between DNS and linear theory, which {gives} consistent predictions for the vertical growth rates of the turbulent cloud and the tilt angles of these structures. Using Lagrangian particle tracking, we distinguish between wave-dominated and turbulence-dominated regions. It is found that the inertial gravity waves emitted by the cloud can transport a substantial fraction of the energy to the neighboring quiescent fluid, which decreases with increasing stratification. This suggests the central role of inertial-gravity waves in energy transfer processes within atmospheres and oceans, particularly around regions of localized turbulence.

The present study elucidates the effects of stratification on localized turbulence under system rotation. First, compared to the purely rotating case, the presence of stratification introduces the zero-frequency PV mode, whose energy distribution barely changes with time. While this mode is not new, our finding is its distinct spatial manifestation within a non-uniformly turbulent background. Second, as the stratification intensifies, the dominant inertial-gravity waves exhibit increased tilting and reduced vertical group velocity. In combination, these phenomena limit the energy emitted by the initial turbulent cloud. {It should be noted that this study focuses on the parameter regime where $Ro<Fr$ (the corresponding $Bu$ ranges from 0 to 1). While the $Ro>Fr$ regime is thought to dominate in the Earth's atmosphere and oceans \citep{riley2000fluid}, the $Ro<Fr$ regime is relevant for several geophysical flows, such as in the deep sea where the influence of stratification is weak but rotation effects are prominent \citep{wingate2011low, heywood2002high, van2005gyroscopic, timmermans2007dynamics, timmermans2010moored}.}% Furthermore, the observed flow structures can be conceptualized as Fourier modes with phase coherence. This provides an alternative perspective: the evolution of these turbulent clouds could be seen as the evolving phase coherence of Fourier modes influenced by both rotation and stratification, which is an avenue for future research.

{There are still many open questions that need to be addressed. Specifically, the potential role of the \emph{Ro}/\emph{Fr} ratio in influencing the flow dynamics and the effect of the Reynolds number warrant further investigation \citep{marino2013inverse, rosenberg2015evidence}. Previous research has shown that the absence of triadic resonance in the parameter range $1/2\le N/(2\Omega)\le2$ is associated with an enhanced inverse cascade \citep{smith2002generation, marino2013inverse}. The particular significance of this range requires further investigation. In addition, the pronounced enhancement of vertical velocity, termed ``vertical drafts'', is a common observation in many DNS studies of stratified flows \citep{feraco2018vertical, marino2022turbulence} and in the oceans \citep{d2007high}. We did not detect such vertical drafts in our simulations, which may be due to the initialization from homogeneous isotropic turbulence rather than rotating stratified turbulence, or the Froude numbers considered are not favorable for such a phenomenon. Given its importance, this phenomenon is worthy of systematic investigation for the development of turbulent clouds in the future. Furthermore, the study of the corresponding Lagrangian statistics is another interesting research direction \citep{buaria2020single}.}

%{While the Lagrangian statistics in rotating stratified turbulence, especially with respect to anisotropy, are certainly intriguing \citep{buaria2020single}, our primary use of Lagrangian particles in this study is to determine the boundaries of turbulence. The study of Lagrangian statistics in our specific scenario, which involves additional inhomogeneity, represents a promising direction for future research.}

%Authors should discuss the results and how they can be interpreted from the perspective of previous studies and of the working hypotheses. The findings and their implications should be discussed in the broadest context possible. Future research directions may also be highlighted.

%%%%%%%%%%%%%%%%%%%%%%%%%%%%%%%%%%%%%%%%%%
% \section{Conclusions}

% This section is not mandatory, but can be added to the manuscript if the discussion is unusually long or complex.

%%%%%%%%%%%%%%%%%%%%%%%%%%%%%%%%%%%%%%%%%%
% \section{Patents}

% This section is not mandatory, but may be added if there are patents resulting from the work reported in this manuscript.

%%%%%%%%%%%%%%%%%%%%%%%%%%%%%%%%%%%%%%%%%%
\vspace{6pt} 

%%%%%%%%%%%%%%%%%%%%%%%%%%%%%%%%%%%%%%%%%%
%% optional
%\supplementary{The following supporting information can be downloaded at:  \linksupplementary{s1}, Figure S1: title; Table S1: title; Video S1: title.}


%%%%%%%%%%%%%%%%%%%%%%%%%%%%%%%%%%%%%%%%%%
\authorcontributions{Conceptualization, T.L., M.W. and S.C.; methodology, T.L.; formal analysis, T.L.; investigation, T.L.; resources, M.W. and S.C.; data curation, T.L.; writing---original draft preparation, T.L.; writing---review and editing, M.W. and S.C.; visualization, T.L.; supervision, M.W. and S.C.; project administration, M.W. and S.C.; funding acquisition, M.W. and S.C. All authors have read and agreed to the published version of the manuscript.}%For research articles with several authors, a short paragraph specifying their individual contributions must be provided. The following statements should be used ``Conceptualization, X.X. and Y.Y.; methodology, X.X.; software, X.X.; validation, X.X., Y.Y. and Z.Z.; formal analysis, X.X.; investigation, X.X.; resources, X.X.; data curation, X.X.; writing---original draft preparation, X.X.; writing---review and editing, X.X.; visualization, X.X.; supervision, X.X.; project administration, X.X.; funding acquisition, Y.Y. All authors have read and agreed to the published version of the manuscript.'', please turn to the  \href{http://img.mdpi.org/data/contributor-role-instruction.pdf}{CRediT taxonomy} for the term explanation. Authorship must be limited to those who have contributed substantially to the work~reported.}

\funding{This research was funded by NSFC Basic Science Center Program, grant number 11988102; NSFC, grant number \TL{12225204}; Department of Science and Technology of Guangdong Province, grant number 2023B1212060001; Shenzhen Science and Technology Innovation Commission, grant number KQTD20180411143441009; and European Research Council (ERC) under the European Union's Horizon 2020 research and innovation programme, grant agreement number 882340.}%and Key Special Project for Introduced Talents Team of Southern Marine Science and Engineering Guangdong Laboratory (Guangzhou), grant number GML2019ZD0103. The APC was funded by [INSERT APC FUNDING SOURCE HERE].}%Please add: ``This research received no external funding'' or ``This research was funded by NAME OF FUNDER grant number XXX.'' and  and ``The APC was funded by XXX''. Check carefully that the details given are accurate and use the standard spelling of funding agency names at \url{https://search.crossref.org/funding}, any errors may affect your future funding.}

\institutionalreview{Not applicable.}%In this section, you should add the Institutional Review Board Statement and approval number, if relevant to your study. You might choose to exclude this statement if the study did not require ethical approval. Please note that the Editorial Office might ask you for further information. Please add “The study was conducted in accordance with the Declaration of Helsinki, and approved by the Institutional Review Board (or Ethics Committee) of NAME OF INSTITUTE (protocol code XXX and date of approval).” for studies involving humans. OR “The animal study protocol was approved by the Institutional Review Board (or Ethics Committee) of NAME OF INSTITUTE (protocol code XXX and date of approval).” for studies involving animals. OR “Ethical review and approval were waived for this study due to REASON (please provide a detailed justification).” OR “Not applicable” for studies not involving humans or animals.}

% \informedconsent{\hl{~~~~~~~~~}}%MDPI: Any research article describing a study involving humans should contain this statement. Please add ``Informed consent was obtained from all subjects involved in the study.'' OR ``Patient consent was waived due to REASON (please provide a detailed justification).'' OR ``Not applicable'' for studies not involving humans. You might also choose to exclude this statement if the study did not involve humans. Written informed consent for publication must be obtained from participating patients who can be identified (including by the patients themselves). Please state ``Written informed consent has been obtained from the patient(s) to publish this paper'' if applicable. % Reply: 9. Thanks. We excluded this statement because it is not applicable

\dataavailability{\hl{The data and code used in this study are available on request from the corresponding author.} \TL{The data and code are not publicly available because we want to understand the specific usage by those requesting access.}}%Please provide specific reason for Data Availability Statement. We encourage all authors of articles published in MDPI journals to share their research data. In this section, please provide details regarding where data supporting reported results can be found, including links to publicly archived datasets analyzed or generated during the study. Where no new data were created, or where data is unavailable due to privacy or ethical re-strictions, a statement is still required. Suggested Data Availability Statements are available in section “MDPI Research Data Policies” at \url{https://www.mdpi.com/ethics}. % Reply: 10. Thanks. Fixed

\acknowledgments{Numerical simulations were supported by the Center for Computational Science and Engineering of Southern University of Science and Technology. T.L. extends gratitude to Vikrant Gupta for proofreading the manuscript. M.W. acknowledges the support from Centers for Mechanical Engineering Research and Education at MIT and SUSTech.}%In this section you can acknowledge any support given which is not covered by the author contribution or funding sections. This may include administrative and technical support, or donations in kind (e.g., materials used for experiments).}

\conflictsofinterest{The authors declare no conflict of interest.}%Declare conflicts of interest or state ``The authors declare no conflict of interest.'' Authors must identify and declare any personal circumstances or interest that may be perceived as inappropriately influencing the representation or interpretation of reported research results. Any role of the funders in the design of the study; in the collection, analyses or interpretation of data; in the writing of the manuscript; or in the decision to publish the results must be declared in this section. If there is no role, please state ``The funders had no role in the design of the study; in the collection, analyses, or interpretation of data; in the writing of the manuscript; or in the decision to publish the~results''.} 

%%%%%%%%%%%%%%%%%%%%%%%%%%%%%%%%%%%%%%%%%%
%% Optional
%\sampleavailability{Samples of the compounds ... are available from the authors.}

%% Only for journal Encyclopedia
%\entrylink{The Link to this entry published on the encyclopedia platform.}

\clearpage
\abbreviations{Abbreviations}{
The following abbreviations are used in this manuscript:\\

\noindent 
\begin{tabular}{@{}ll}
DNS & Direct numerical simulation\\
PV &  Potential vorticity
% MDPI & Multidisciplinary Digital Publishing Institute\\
% DOAJ & Directory of open access journals\\
% TLA & Three letter acronym\\
% LD & Linear dichroism
\end{tabular}
}

%%%%%%%%%%%%%%%%%%%%%%%%%%%%%%%%%%%%%%%%%%
%% Optional
% \appendixtitles{no} % Leave argument "no" if all appendix headings stay EMPTY (then no dot is printed after "Appendix A"). If the appendix sections contain a heading then change the argument to "yes".
% \appendixstart
% \appendix
% \section[\appendixname~\thesection]{}
% \subsection[\appendixname~\thesubsection]{}
% The appendix is an optional section that can contain details and data supplemental to the main text---for example, explanations of experimental details that would disrupt the flow of the main text but nonetheless remain crucial to understanding and reproducing the research shown; figures of replicates for experiments of which representative data are shown in the main text can be added here if brief, or as Supplementary Data. Mathematical proofs of results not central to the paper can be added as an appendix.

% \begin{table}[H] 
% \caption{This is a table caption.\label{tab5}}
% \newcolumntype{C}{>{\centering\arraybackslash}X}
% \begin{tabularx}{\textwidth}{CCC}
% \toprule
% \textbf{Title 1}	& \textbf{Title 2}	& \textbf{Title 3}\\
% \midrule
% Entry 1		& Data			& Data\\
% Entry 2		& Data			& Data\\
% \bottomrule
% \end{tabularx}
% \end{table}

% \section[\appendixname~\thesection]{}
% All appendix sections must be cited in the main text. In the appendices, Figures, Tables, etc. should be labeled, starting with ``A''---e.g., Figure A1, Figure A2, etc.

%%%%%%%%%%%%%%%%%%%%%%%%%%%%%%%%%%%%%%%%%%
\begin{adjustwidth}{-\extralength}{0cm}

\reftitle{References}

\PublishersNote{}
\end{adjustwidth}
\end{document}